\documentclass{article}

\usepackage[T1]{fontenc}
\usepackage{tgtermes}

\usepackage{amssymb}
\newcommand{\mathbold}[1]{\ensuremath{\boldsymbol{\mathbf{#1}}}}
\usepackage{amsmath}

\usepackage[
  paper  = letterpaper,
  left   = 1.50in,
  right  = 1.50in,
  top    = 1.0in,
  bottom = 1.0in,
  ]{geometry}
\usepackage[english]{babel}
\usepackage[parfill]{parskip}
\usepackage{afterpage}
\usepackage{enumitem}
\usepackage{framed}
\usepackage{xspace}

\usepackage[usenames,dvipsnames]{xcolor}
\definecolor{shadecolor}{gray}{0.95}

\usepackage{graphicx}
\usepackage[labelfont=bf]{caption}
\usepackage[format=hang]{subcaption}

\usepackage{booktabs, array}

\usepackage{natbib}

\usepackage{listings}

\definecolor{strings}{rgb}{.624,.251,.259}
\definecolor{keywords}{rgb}{.224,.451,.686}
\definecolor{comment}{rgb}{.322,.451,.322}

\lstdefinelanguage{python}{
  morekeywords={from, import, as, for, in, while, def, return, =, +,
  -, /, *, lambda},
  keywords=[2]{build_toy_dataset, neural_network, __init__},
  keywords=[3]{Normal, Bernoulli, Beta, Categorical, Dirichlet,
  Exponential, MultivariateNormalFull, RandomVariable, Distribution,
  DirichletProcess, Empirical, PointMass, Gamma,
  MAP, Inference, KLqp, HMC, SGLD, KLpq,
  VariationalInference, MonteCarlo, ConjugateInference, GANInference,
  rnn_cell, dirichlet_process, cond, body,
  evaluate, ppc, copy, dot, get_session},
  morecomment=[l]{\#},
  morecomment=[s]{"""}{"""},
  morestring=[b]',
  morestring=[b]",
  alsoletter={<>=-+/*},
  sensitive=true
}

\renewcommand{\texttt}[1]{\lstinline[basicstyle=\fontsize{8pt}{8.25pt}\selectfont\ttfamily]{#1}}

\usepackage[colorlinks,linktoc=all, bookmarks=false]{hyperref}
\usepackage[all]{hypcap}
\hypersetup{citecolor=MidnightBlue}
\hypersetup{linkcolor=MidnightBlue}
\hypersetup{urlcolor=MidnightBlue}

\usepackage[acronym,smallcaps,nowarn]{glossaries}
\glsdisablehyper{}

\newacronym{VI}{vi}{variational inference}
\newacronym{KL}{kl}{Kullback-Leibler}
\newacronym{ELBO}{elbo}{\emph{evidence lower bound}}
\newacronym{MCMC}{mcmc}{Markov chain Monte Carlo}
\newacronym{PPC}{ppc}{posterior predictive check}

\newcommand{\g}{\,|\,}

\newcommand{\mba}{\mathbold{a}}

\newcommand{\mbp}{\mathbold{p}}

\newcommand{\mbx}{\mathbold{x}}

\newcommand{\mbz}{\mathbold{z}}

\usepackage{tikz}
\usetikzlibrary{bayesnet}

\pgfdeclarelayer{edgelayer}
\pgfdeclarelayer{nodelayer}
\pgfsetlayers{edgelayer,nodelayer,main}

\definecolor{hexcolor0xbfbfbf}{rgb}{0.749,0.749,0.749}

\tikzset{>=latex}
\tikzstyle{none}   = [inner sep=0pt]
\tikzstyle{line}   = [-,
                      thick,
                      shorten <=1pt,
                      shorten >=1pt]
\tikzstyle{arrow}  = [->,
                      thick,
                      shorten <=1pt,
                      shorten >=1pt]
\tikzstyle{ardash} = [dashed,
                      ->,
                      thick,
                      shorten <=1pt,
                      shorten >=1pt]

\tikzstyle{empty}=[
                   circle,
                   opacity=0.0,
                   text opacity=1.0,
                   inner sep=0pt
                  ]

\tikzstyle{box}=[
                 rectangle,
                 fill=White,
                 thick,
                 draw=Black,
                 inner sep=7pt
                ]

\tikzstyle{filled}=[
                    circle,
                    thick,
                    fill=hexcolor0xbfbfbf,
                    draw=Black
                   ]

\tikzstyle{hollow}=[
                    circle,
                    thick,
                    fill=White,
                    draw=Black
                   ]

\tikzstyle{param}=[
                   rectangle,
                   fill=Black,
                   draw=Black,
                   inner sep=0pt,
                   minimum width=4pt,
                   minimum height=4pt
                  ]

\tikzstyle{paramhollow}=[
                         rectangle,
                         thick,
                         fill=White,
                         draw=Black,
                         inner sep=0pt,
                         minimum
                         width=4pt,
                         minimum height=4pt
                        ]

\usepackage{pgfplots}                               
\pgfplotsset{compat=newest}
\pgfplotsset{plot coordinates/math parser=false}

\title{
Edward: A library for probabilistic modeling,\break inference, and criticism
}

\author{
Dustin Tran, Alp Kucukelbir, Adji B.~Dieng,\\
Maja Rudolph, Dawen Liang, and David M.~Blei\\\\
Columbia University
}

\begin{document}

\hypersetup{pageanchor=false}
\begin{titlepage}
\maketitle
\begin{abstract}
  Probabilistic modeling is a powerful approach for analyzing
  empirical information.
  We describe \emph{Edward}, a library for probabilistic modeling.
  Edward's design reflects an iterative process pioneered by George
  Box: build a model of a phenomenon, make inferences about the model given
  data, and criticize the model's fit to the data. Edward supports a broad class
  of probabilistic models, efficient algorithms for inference, and many
  techniques for model criticism. The library builds on top of
  TensorFlow to support distributed training and hardware such as
  GPUs.
  Edward enables the development of complex probabilistic
  models and their algorithms at a massive scale.
\end{abstract}

\footnote{Details in this paper describe Edward version 1.2.1,
released Jan 30, 2017.}

\setcounter{secnumdepth}{2}
\setcounter{tocdepth}{2}
\tableofcontents

\thispagestyle{empty}
\end{titlepage}
\hypersetup{pageanchor=true}

\glsresetall{}

\clearpage

\section{Introduction}
\label{sec:intro}

Probabilistic modeling is a powerful approach for analyzing empirical
information \citep{tukey1962future,newell1976computer,box1976science}.
Probabilistic models are essential to fields related to its
methodology, such as
statistics \citep{friedman2001elements,gelman2013bayesian} and machine
learning \citep{murphy2012machine,goodfellow2016deep}, as well as
fields related to its application, such as computational
biology \citep{friedman2000using}, computational
neuroscience \citep{dayan2001theoretical}, cognitive
science \citep{tenenbaum2011grow}, information
theory \citep{mackay2003information}, and natural language
processing \citep{manning1999foundations}.

Software systems for probabilistic modeling provide new and faster
ways of experimentation. This enables research advances in
probabilistic modeling that could not have been completed before.

As an example of such software systems, we point to early work in
artificial intelligence.
Expert systems were designed from human expertise, which in
turn enabled larger reasoning steps according to existing
knowledge \citep{buchanan1969heuristic,minsky1975framework}.  With
connectionist models, the design focused on neuron-like processing
units, which learn from experience;
this drove new applications of artificial intelligence
\citep{hopfield1982neural,rumelhart1988parallel}.

As another example, we point to early work in
statistical computing, where interest grew broadly out of efficient
computation for problems in statistical analysis. The S language,
developed by John Chambers and colleagues at Bell
Laboratories \citep{becker1984s,chambers1992statistical}, focused on
an interactive environment for data analysis, with simple yet rich
syntax to quickly turn ideas into software. It is a
predecessor to the R language \citep{ihaka1996r}.  More targeted environments
such as BUGS \citep{spiegelhalter1995bugs},
which focuses on Bayesian analysis of statistical models, helped launch
the emerging field of probabilistic programming.

We are motivated to build on these early works in probabilistic
systems---where in modern applications, new challenges arise in their
design and implementation.  We highlight two challenges.
First, statistics and machine learning have made significant advances
in the methodology of probabilistic models and their inference (e.g.,
\citet{hoffman2013stochastic,ranganath2014black,rezende2014stochastic}).
For software systems to enable fast experimentation, we require rich
abstractions that can capture these advances: it must encompass both a
broad class of probabilistic models and a broad class of algorithms
for their efficient inference.
Second, researchers are increasingly motivated to employ complex
probabilistic models and at an unprecedented scale of massive
data \citep{bengio2013representation,ghahramani2015probabilistic,lake2016building}.
Thus we require an efficient computing environment that supports
distributed training and integration of hardware such as (multiple)
GPUs.

We present \emph{Edward}, a probabilistic modeling library named after the
statistician George Edward Pelham Box. Edward is built around an iterative
process for probabilistic modeling, pioneered by Box and his collaborators
\citep{box1962useful,box1965experimental,box1967discrimination,box1976science,
box1980sampling}. The process is as follows: given data from some unknown
phenomena, first, formulate a model of the phenomena; second, use an algorithm
to infer the model's hidden structure, thus reasoning about the phenomena;
third, criticize how well the model captures the data's generative process. As
we criticize our model's fit to the data, we revise components of the model and
repeat to form an iterative
loop \citep{box1976science,blei2014build,gelman2013bayesian}.

Edward builds infrastructure to enable this loop:
\begin{enumerate}
\item
For \emph{modeling}, Edward provides a language of random variables to
construct a broad class of models: directed graphical
models \citep{pearl1988probabilistic}, stochastic neural
networks \citep{neal1990learning}, and programs with stochastic
control flow \citep{goodman2012church}.
\item
For \emph{inference}, Edward provides algorithms such as stochastic
and black box variational
inference \citep{hoffman2013stochastic,ranganath2014black},
Hamiltonian Monte Carlo \citep{neal1993probabilistic}, and stochastic
gradient Langevin dynamics \citep{welling2011bayesian}. Edward also
provides infrastructure to make it easy to develop new algorithms.
\item
For \emph{criticism}, Edward provides methods from scoring
rules \citep{winkler1996scoring} and predictive
checks \citep{box1980sampling,rubin1984bayesianly}.
\end{enumerate}
Edward is built on top of TensorFlow, a library for numerical computing
using data flow graphs \citep{abadi2016tensorflow}. TensorFlow enables
Edward to speed up computation with hardware such as GPUs, to scale
up computation with distributed training, and to simplify engineering
effort with automatic differentiation.

In Section\nobreakspace \ref {sec:getting-started},
we demonstrate Edward with an example.
In Section\nobreakspace \ref {sec:design},
we describe the design of Edward.
In Section\nobreakspace \ref {sec:examples},
we provide examples of how standard tasks in statistics and machine learning can be solved with Edward.

\subsection*{Related work}

\textbf{Probabilistic programming.}
There has been much work on programming languages which specify
broad classes of probabilistic models, or probabilistic
programs. Recent works include
Church \citep{goodman2012church},
Venture \citep{mansinghka2014venture},
Anglican \citep{wood2015new},
Stan \citep{carpenter2016stan}, and
WebPPL \citep{goodman2014design}.
The most important distinction in Edward stems from motivation. We
are interested in deploying probabilistic models to many real
world applications, ranging from the size of data and data structure,
such as large text corpora or many brief audio signals, to the size of
model and class of models, such as small nonparametric models or deep
generative models. Thus Edward is built with fast computation in
mind.

\textbf{Black box inference.}
Black box algorithms are typically based on Monte Carlo methods, and
make very few assumptions about the
model \citep{metropolis1949monte,hastings1970monte,geman1984stochastic}.
Our motivation as outlined
above presents a new set of challenges in both inference research and
software design. As a first consequence, we focus on variational
inference \citep{hinton1993keeping,waterhouse1996bayesian,jordan1999introduction}.
As a second consequence, we encourage active research on
inference by providing a class hierarchy of inference algorithms. As a
third consequence, our inference algorithms aim to take advantage of
as much structure as possible from the model. Edward supports all
types of inference, whether they
be black box or model-specific \citep{dempster1977em,hoffman2013stochastic}.

\textbf{Computational frameworks.}
There are many computational frameworks, primarily built for deep
learning: as of this date, this includes
TensorFlow \citep{abadi2016tensorflow},
Theano \citep{alrfou2016theano},
Torch \citep{collobert2011torch7},
neon \citep{nervana2014neon}, and
the Stan Math Library \citep{carpenter2015stan}.
These are incredible tools which Edward employs as a backend. In terms
of abstraction, Edward sits one level higher.

\textbf{High-level deep learning libraries.}
Neural network libraries such as
Keras \citep{chollet2015keras} and
Lasagne \citep{lasagne}
are at a similar abstraction level as Edward. However both are primarily
interested in parameterizing complicated functions for supervised
learning on large datasets. We are interested in probabilistic models
which apply to a wide array of learning tasks. These tasks may have both
complicated likelihood and complicated priors (neural networks are an
option but not a necessity). Therefore our goals are orthogonal and in
fact mutually beneficial to each other. For example, we use Keras'
abstraction as a way to easily specify models parameterized by deep
neural networks.

\clearpage
\section{Getting Started}
\label{sec:getting-started}

Probabilistic modeling in Edward uses a simple language of
random variables.
Here we will show a Bayesian neural network. It is a neural network
with a prior distribution on its weights.

First, simulate a toy dataset of 50 observations with a cosine relationship.

\begin{lstlisting}
import numpy as np

x_train = np.linspace(-3, 3, num=50)
y_train = np.cos(x_train) + np.random.normal(0, 0.1, size=50)
x_train = x_train.astype(np.float32).reshape((50, 1))
y_train = y_train.astype(np.float32).reshape((50, 1))
\end{lstlisting}

\begin{figure}[h]
\centering
\includegraphics[width=3.5in]{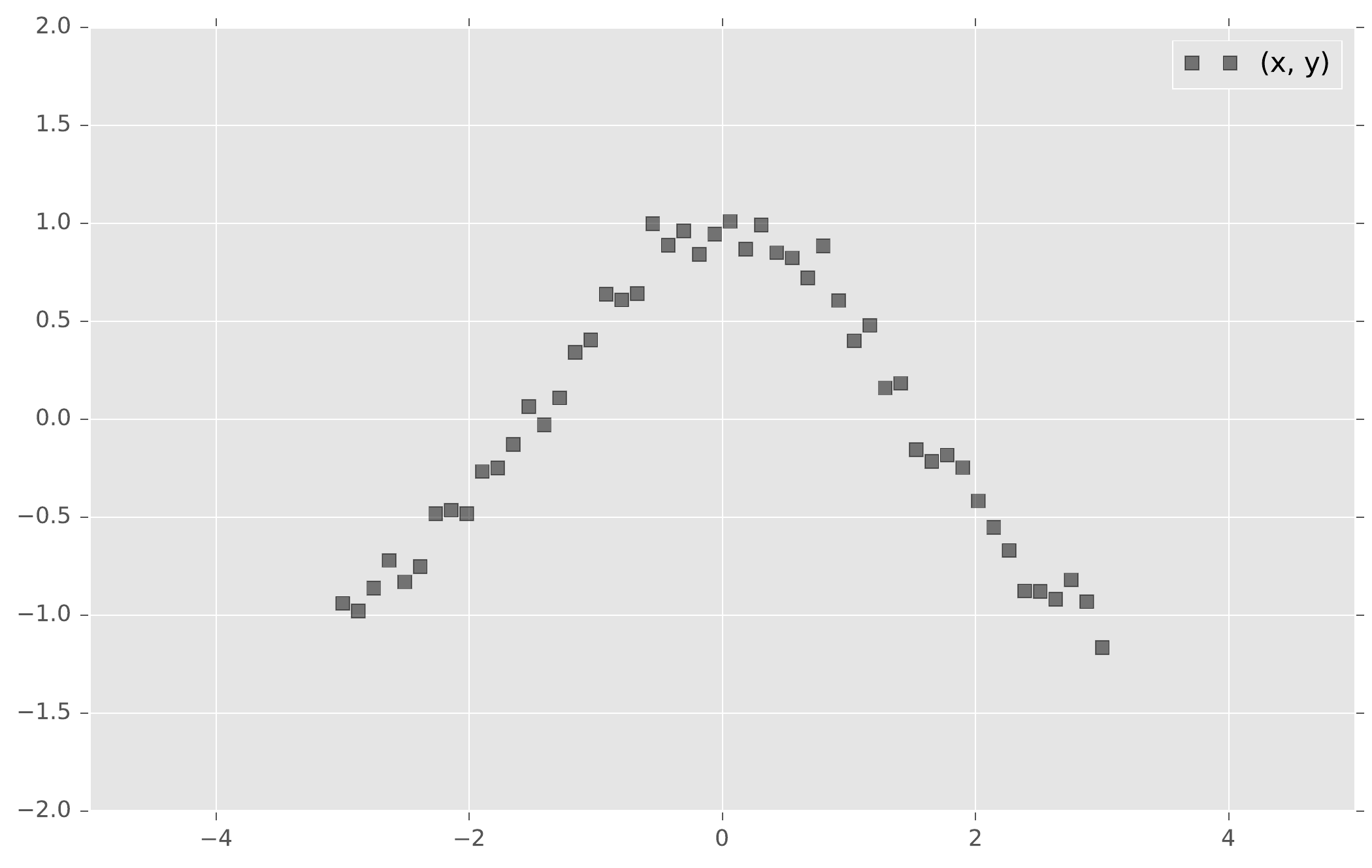}
\caption{Simulated data with a cosine relationship between $x$ and $y$.}
\end{figure}

Next, define a two-layer Bayesian neural network. Here, we
define the neural network manually with \texttt{tanh} nonlinearities.

\begin{lstlisting}[language=Python]
import tensorflow as tf
from edward.models import Normal

W_0 = Normal(mu=tf.zeros([1, 2]), sigma=tf.ones([1, 2]))
W_1 = Normal(mu=tf.zeros([2, 1]), sigma=tf.ones([2, 1]))
b_0 = Normal(mu=tf.zeros(2), sigma=tf.ones(2))
b_1 = Normal(mu=tf.zeros(1), sigma=tf.ones(1))

x = x_train
y = Normal(mu=tf.matmul(tf.tanh(tf.matmul(x, W_0) + b_0), W_1) + b_1,
           sigma=0.1)
\end{lstlisting}

Next, make inferences about the model from data. We will use variational
inference. Specify a normal approximation over the weights and biases.

\begin{lstlisting}[language=Python]
qW_0 = Normal(mu=tf.Variable(tf.zeros([1, 2])),
              sigma=tf.nn.softplus(tf.Variable(tf.zeros([1, 2]))))
qW_1 = Normal(mu=tf.Variable(tf.zeros([2, 1])),
              sigma=tf.nn.softplus(tf.Variable(tf.zeros([2, 1]))))
qb_0 = Normal(mu=tf.Variable(tf.zeros(2)),
              sigma=tf.nn.softplus(tf.Variable(tf.zeros(2))))
qb_1 = Normal(mu=tf.Variable(tf.zeros(1)),
              sigma=tf.nn.softplus(tf.Variable(tf.zeros(1))))
\end{lstlisting}

Defining \texttt{tf.Variable} allows the variational factors'
parameters to vary. They are all initialized at 0. The standard
deviation parameters are constrained to be greater than zero according
to a
softplus transformation\footnote{
The softplus function is defined as $\textrm{softplus}(x) =
\log(1+\exp(x))$.}.

Now, run variational inference with the
Kullback-Leibler divergence
in order to infer the model's latent variables given data.
We specify \texttt{1000} iterations.

\begin{lstlisting}[language=Python]
import edward as ed

inference = ed.KLqp({W_0: qW_0, b_0: qb_0,
                     W_1: qW_1, b_1: qb_1}, data={y: y_train})
inference.run(n_iter=1000)
\end{lstlisting}

Finally, criticize the model fit. Bayesian neural networks define a distribution
over neural networks, so we can perform a graphical check. Draw neural networks
from the inferred model and visualize how well it fits the data.

\begin{figure}[h]
\centering
\includegraphics[width=3.5in]{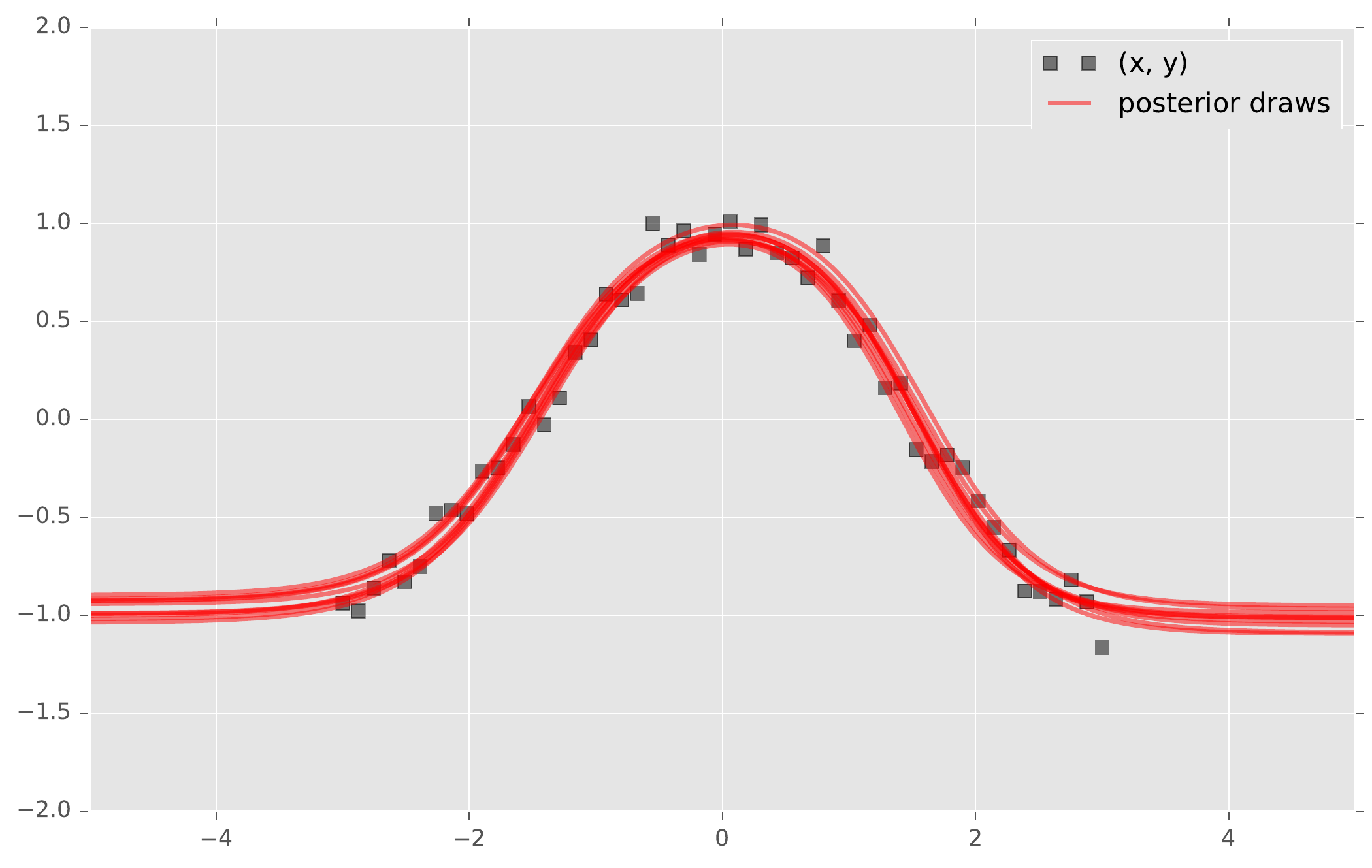}
\caption{Posterior draws from the inferred Bayesian neural network.}
\end{figure}

The model has captured the cosine relationship between $x$ and $y$
in the observed domain.

\clearpage
\section{Design}
\label{sec:design}

Edward's design reflects the building blocks for probabilistic
modeling. It defines interchangeable components, enabling rapid
experimentation and research with probabilistic models.

Edward is named after the innovative statistician
George Edward Pelham Box. Edward follows Box's philosophy of statistics and
machine learning \citep{box1976science}.

First gather data from some real-world phenomena. Then cycle through
Box's loop \citep{blei2014build}.

\begin{enumerate}
\item Build a probabilistic model of the phenomena.
\item Reason about the phenomena given model and data.
\item Criticize the model, revise and repeat.
\end{enumerate}

\begin{figure}[htb]
\centering
\begin{tikzpicture}
	\begin{pgfonlayer}{nodelayer}
		\node [style=box] (0) at (-2.5, 0) {Model};
		\node [style=box] (1) at (0, 0) {Infer};
		\node [style=box, fill=MidnightBlue, draw=white] (2) at (0, 1.5)
		{\color{white}Data};
		\node [style=box] (3) at (2.75, 0) {Criticize};
	\end{pgfonlayer}
	\begin{pgfonlayer}{edgelayer}
		\draw [style=arrow] (0) to (1);
		\draw [style=arrow] (2) to (1);
		\draw [style=arrow] (1) to (3);
		\draw [style=arrow, bend left=75, looseness=1.25] (3) to (0);
	\end{pgfonlayer}
\end{tikzpicture}
\caption{Box's loop.}
\end{figure}
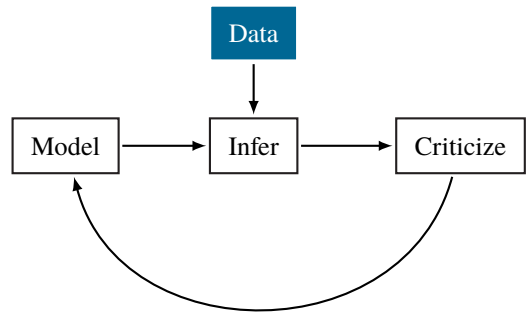

Here's a toy example. A child flips a coin ten times, with the set of outcomes
being

\texttt{{[}0,\ 1,\ 0,\ 0,\ 0,\ 0,\ 0,\ 0,\ 0,\ 1{]}},

where \texttt{0}
denotes tails and \texttt{1} denotes heads. She is interested in the
probability that the coin lands heads. To analyze this, she first
builds a model: suppose she assumes the coin flips are independent and
land heads with the same probability. Second, she reasons about the
phenomenon: she infers the model's hidden structure given data.
Finally, she criticizes the model: she analyzes whether her model
captures the real-world phenomenon of coin flips. If it doesn't, then
she may revise the model and repeat.

We describe modules enabling this analysis.

\subsection{Data}

Data defines a set of observations. There are three ways
to read data in Edward.

\textbf{Preloaded data.}
A constant or variable in the TensorFlow graph holds all the data.
This setting is the fastest to work with and is recommended if the
data fits in memory.

Represent the data as NumPy arrays or TensorFlow tensors.

\begin{lstlisting}[language=Python]
x_data = np.array([0, 1, 0, 0, 0, 0, 0, 0, 0, 1])
x_data = tf.constant([0, 1, 0, 0, 0, 0, 0, 0, 0, 1])
\end{lstlisting}

During inference, we store them in TensorFlow variables internally to
prevent copying data more than once in memory.

\textbf{Feeding.}
Manual code provides the data when running each step of inference.
This setting provides the most fine control which is useful for
experimentation.

Represent the data as
TensorFlow placeholders,
which are nodes in the graph that are fed at runtime.

\begin{lstlisting}[language=Python]
x_data = tf.placeholder(tf.float32, [100, 25])  # placeholder of shape (100, 25)
\end{lstlisting}

During inference, the user must manually feed the placeholders. At each
step, call \texttt{inference.update()} while
passing in a \texttt{feed\_dict} dictionary
which binds placeholders to realized values as an argument.
If the values do not change over inference updates, one can also bind
the placeholder to values within the \texttt{data} argument when
first constructing inference.

\textbf{Reading from files.}
An input pipeline reads the data from files at the beginning of a
TensorFlow graph. This setting is recommended if the data does not
fit in memory.

\begin{lstlisting}[language=Python]
filename_queue = tf.train.string_input_producer(...)
reader = tf.SomeReader()
...
\end{lstlisting}

Represent the data as TensorFlow tensors, where the tensors are the
output of data readers. During inference, each update will be
automatically evaluated over new batch tensors represented through
the data readers.

\subsection{Models}

A probabilistic model is a joint distribution $p(\mathbf{x},
\mathbf{z})$ of data $\mathbf{x}$ and latent variables $\mathbf{z}$.

In Edward, we specify models using a simple language of random variables.
A random variable $\mathbf{x}$ is an object parameterized by
tensors $\theta^*$, where
the number of random variables in one object is determined by
the dimensions of its parameters.

\begin{lstlisting}[language=Python]
from edward.models import Normal, Exponential

# univariate normal
Normal(mu=tf.constant(0.0), sigma=tf.constant(1.0))
# vector of 5 univariate normals
Normal(mu=tf.zeros(5), sigma=tf.ones(5))
# 2 x 3 matrix of Exponentials
Exponential(lam=tf.ones([2, 3]))
\end{lstlisting}

For multivariate distributions, the multivariate dimension is the
innermost (right-most) dimension of the parameters.

\begin{lstlisting}[language=Python]
from edward.models import Dirichlet, MultivariateNormalFull

# K-dimensional Dirichlet
Dirichlet(alpha=tf.constant([0.1] * K)
# vector of 5 K-dimensional multivariate normals
MultivariateNormalFull(mu=tf.zeros([5, K]), sigma=...)
# 2 x 5 matrix of K-dimensional multivariate normals
MultivariateNormalFull(mu=tf.zeros([2, 5, K]), sigma=...)
\end{lstlisting}

Random variables are equipped with methods such as
\texttt{log_prob()}, $\log p(\mathbf{x}\mid\theta^*)$,
\texttt{mean()}, $\mathbb{E}_{p(\mathbf{x}\mid\theta^*)}[\mathbf{x}]$,
and \texttt{sample()}, $\mathbf{x}^*\sim p(\mathbf{x}\mid\theta^*)$.
Further, each random variable is associated to a tensor $\mathbf{x}^*$ in the
computational graph, which represents a single sample $\mathbf{x}^*\sim
p(\mathbf{x}\mid\theta^*)$.

This makes it easy to parameterize random variables with complex
deterministic structure, such as with deep neural networks, a diverse
set of math operations, and compatibility with third party libraries
which also build on TensorFlow.
The design also enables compositions of random variables
to capture complex stochastic structure.
They operate on $\mathbf{x}^*$.

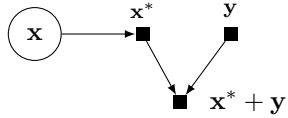
\begin{figure}[!htb]
\centering
\begin{tikzpicture}[x=1.7cm,y=1.8cm,scale=0.9]

  \node[latent] (x) {$\mathbf{x}$};
  \factor[right=of x, xshift=0.3cm] {xstar} {$\mathbf{x}^*$} {} {};
  \factor[right=of xstar, xshift=0.3cm] {y} {$\mathbf{y}$} {} {};

  \factor[below=of xstar, xshift=0.5cm] {plus} {} {} {};
  \node at (plus) [right=0.175cm of plus] {$\mathbf{x}^*+\mathbf{y}$};

  \edge{x}{xstar};
  \edge{xstar}{plus};
  \edge{y}{plus};

\end{tikzpicture}
\caption{Random variables can be combined with other TensorFlow ops.}
\end{figure}

\begin{lstlisting}[language=Python]
from edward.models import Normal

x = Normal(mu=tf.zeros(10), sigma=tf.ones(10))
y = tf.constant(5.0)
x + y, x - y, x * y, x / y
tf.tanh(x * y)
tf.gather(x, 2)  # 3rd normal rv in the vector
\end{lstlisting}

Below we describe how to build models by composing random variables.

For a list of random variables supported in Edward, see the
{TensorFlow distributions API}.
\footnote{\url{https://www.tensorflow.org/versions/master/api_docs/python/contrib.distributions.html}}
Edward random variables build on top of them, inheriting the same
arguments and class methods. Additional methods are also available,
detailed in Edward's API.

\subsection*{Composing Random Variables}

Core to Edward's design is compositionality. Compositionality enables
fine control of modeling, where models are represented as a collection
of random variables.

We outline how to write popular classes of models using Edward:
directed graphical models, neural networks, Bayesian nonparametrics,
and probabilistic programs.

\subsubsection{Directed Graphical Models}

Graphical models are a rich formalism for specifying probability
distributions \citep{koller2009probabilistic}.
In Edward, directed edges in a graphical model are implicitly defined
when random variables are composed with one another. We illustrate
with a Beta-Bernoulli model,
\begin{equation*}
p(\mathbf{x}, \theta) =
\text{Beta}(\theta\mid 1, 1)
\prod_{n=1}^{50} \text{Bernoulli}(x_n\mid \theta),
\end{equation*}
where $\theta$ is a latent probability shared across the 50 data
points $\mathbf{x}\in\{0,1\}^{50}$.

\begin{lstlisting}[language=python]
from edward.models import Bernoulli, Beta

theta = Beta(a=1.0, b=1.0)
x = Bernoulli(p=tf.ones(50) * theta)
\end{lstlisting}

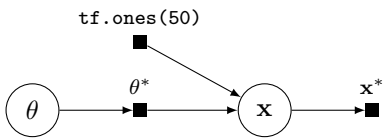
\begin{figure}[!htb]
\centering
\begin{tikzpicture}[x=1.7cm,y=1.8cm,scale=0.9]

  \node[latent] (theta) {$\theta$};
  \factor[right=of theta, xshift=0.3cm] {thetastar} {$\theta^*$} {} {};

  \factor[above=of thetastar] {n} {\texttt{tf.ones(50)}} {} {};
  \node[latent, right=of thetastar, xshift=-0.5cm] (x) {$\mbx$};
  \factor[right=of x, xshift=0.3cm] {xstar} {$\mbx^*$} {} {};

  \edge{theta}{thetastar};
  \edge{thetastar}{x};
  \edge{n}{x};
  \edge{x}{xstar};

\end{tikzpicture}
\caption{Computational graph for a Beta-Bernoulli program.}
\end{figure}

The random variable \texttt{x} ($\mathbf{x}$) is 50-dimensional,
parameterized by the random tensor $\theta^*$. Fetching the object
\texttt{x.value()} ($\mathbf{x}^*$) from session runs the graph: it simulates from
the generative process and outputs a binary vector of $50$ elements.

With computational graphs, it is also natural to build mutable states
within the probabilistic program. As a typical use of computational
graphs, such states can define model parameters, that is, parameters
that we will always compute point estimates for and not be uncertain
about. In TensorFlow, this is given by a \texttt{tf.Variable}.

\begin{lstlisting}[language=python]
from edward.models import Bernoulli

theta = tf.Variable(0.0)
x = Bernoulli(p=tf.ones(50) * tf.sigmoid(theta))
\end{lstlisting}

Another use case of mutable states is for building discriminative
models $p(\mathbf{y}\mid\mathbf{x})$, where $\mathbf{x}$ are features
that are input as training or test data. The program can be written
independent of the data, using a mutable state
(\texttt{tf.placeholder}) for $\mathbf{x}$ in its graph. During
training and testing, we feed the placeholder the appropriate values.

\subsubsection{Neural Networks}

As Edward uses TensorFlow, it is easy to construct neural networks for
probabilistic modeling \citep{rumelhart1988parallel}.
For example, one can specify stochastic neural networks
\citep{neal1990learning}.

High-level libraries such as
{Keras}\footnote{\url{http://keras.io}} and
{TensorFlow Slim}\footnote{\url{https://github.com/tensorflow/tensorflow/tree/master/tensorflow/contrib/slim}}
can be used to easily construct deep neural networks.
We illustrate this with a deep generative model over binary data
$\{\mathbf{x}_n\}\in\{0,1\}^{N\times 28*28}$.

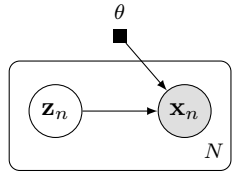
\begin{figure}[!htb]
\centering
\begin{tikzpicture}

  \node[latent] (z) {$\mbz_n$};
  \node[obs, right=of z] (x) {$\mbx_n$};

  \factor[empty, right=of z] {h} {} {} {};
  \factor[right=of z, yshift=1.0cm] {theta} {$\theta$} {} {};

  \edge{z}{x};
  \edge{theta}{x};

  \plate[inner sep=0.25cm, yshift=0.05cm,
    label={[xshift=-14pt,yshift=14pt]south east:$N$}] {plate1} {
    (z)(x)
  } {};

\end{tikzpicture}
\caption{Graphical representation of a deep generative model.}
\end{figure}

The model specifies a generative process where for each
$n=1,\ldots,N$,
\begin{align*}
\mathbf{z}_n &\sim \text{Normal}(\mathbf{z}_n \mid \mathbf{0}, \mathbf{I}), \\
\mathbf{x}_n\mid \mathbf{z}_n &\sim \text{Bernoulli}(\mathbf{x}_n\mid
p=\mathrm{NN}(\mathbf{z}_n; \mathbf{\theta})).
\end{align*}
The latent space is $\mathbf{z}_n\in\mathbb{R}^d$ and the
likelihood is parameterized by a neural network $\mathrm{NN}$ with
parameters $\theta$. We will use a two-layer neural network with a
fully connected hidden layer of 256 units (with ReLU activation) and
whose output is $28*28$-dimensional. The output will be unconstrained,
parameterizing the logits of the Bernoulli likelihood.

With TensorFlow Slim, we write this model as follows:

\begin{lstlisting}[language=python]
from edward.models import Bernoulli, Normal
from tensorflow.contrib import slim

z = Normal(mu=tf.zeros([N, d]), sigma=tf.ones([N, d]))
h = slim.fully_connected(z, 256)
x = Bernoulli(logits=slim.fully_connected(h, 28 * 28, activation_fn=None))
\end{lstlisting}

With Keras, we write this model as follows:

\begin{lstlisting}[language=python]
from edward.models import Bernoulli, Normal
from keras.layers import Dense

z = Normal(mu=tf.zeros([N, d]), sigma=tf.ones([N, d]))
h = Dense(256, activation='relu')(z.value())
x = Bernoulli(logits=Dense(28 * 28)(h))
\end{lstlisting}

Keras and TensorFlow Slim automatically manage TensorFlow variables, which
serve as parameters of the high-level neural network layers. This
saves the trouble of having to manage them manually. However, note
that neural network parameters defined this way always serve as model
parameters. That is, the parameters are not exposed to the user so we
cannot be Bayesian about them with prior distributions.

\subsubsection{Bayesian Nonparametrics}

Bayesian nonparametrics enable rich probability models by working over
an infinite-dimensional parameter space \citep{hjort2010bayesian}.
Edward supports the two typical approaches to handling these models:
collapsing the infinite-dimensional space and lazily defining the
infinite-dimensional space.

For the collapsed approach, see the
Gaussian process classification
tutorial as an example. We specify distributions over the function
evaluations of the Gaussian process, and the Gaussian process is
implicitly marginalized out. This approach is also useful for Poisson
process models.

To work directly on the infinite-dimensional space, one can leverage
random variables with
control flow operations
in TensorFlow. At runtime, the control flow will lazily define any
parameters in the space necessary in order to generate samples. As an
example, we use a while loop to define a
Dirichlet process according to its stick breaking representation.

\subsubsection{Probabilistic Programs}

Probabilistic programs greatly expand the scope of probabilistic
models \citep{goodman2012church}.
Formally, Edward is a Turing-complete probabilistic programming
language. This means that Edward can represent any computable
probability distribution.

\begin{figure}[!htb]
\centering
\begin{tikzpicture}[x=1.7cm,y=1.8cm,scale=0.9]

  \node[latent] (p) {$\mbp$};
  \factor[right=of p, xshift=0.3cm] {pstar} {$\mbp^*$} {} {};

  \factor[above=of pstar] {n} {} {} {};
  \factor[empty, right=of n, yshift=0.1cm] {nn} {\texttt{tf.while_loop(...)}} {} {};
  \factor[left=of n, xshift=-0.5cm] {astar} {$\mba^*$} {} {};
  \node[latent, left=of astar, xshift=0.5cm] (a) {$\mba$};
  \node[latent, right=of pstar, xshift=-0.5cm] (x) {$\mbx$};
  \factor[right=of x, xshift=0.3cm] {xstar} {$\mbx^*$} {} {};

  \edge{p}{pstar};
  \edge{pstar}{x};
  \edge{n}{x};
  \edge{x}{xstar};
  \edge{a}{astar};
  \edge{astar}{n};

\end{tikzpicture}
\caption{Computational graph for a probabilistic program with stochastic control flow.}
\end{figure}
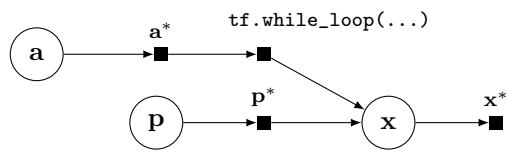

Random variables can be composed with control flow operations,
enabling probabilistic programs with stochastic control flow.
Stochastic control flow defines dynamic conditional dependencies,
known in the literature as contingent or existential dependencies
\citep{mansinghka2014venture,wu2016swift}.
See above, where $\mathbf{x}$ may or may not depend on $\mathbf{a}$
for a given execution.

Stochastic control flow produces difficulties for algorithms that
leverage the graph structure; the relationship of conditional
dependencies changes across execution traces.
Importantly, the computational graph provides an elegant way of
teasing out static conditional dependence structure ($\mathbf{p}$)
from dynamic dependence structure ($\mathbf{a})$. We can perform
model parallelism over the static structure with GPUs and batch
training, and use generic computations to handle the dynamic
structure.

\subsection*{Developing Custom Random Variables}

Oftentimes we'd like to implement our own random variables.
To do so, write a class that inherits
the \texttt{RandomVariable} class in \texttt{edward.models} and
the \texttt{Distribution} class in \texttt{tf.contrib.distributions} (in that
order). A template is provided below.

\begin{lstlisting}[language=Python]
from edward.models import RandomVariable
from tensorflow.contrib.distributions import Distribution

class CustomRandomVariable(RandomVariable, Distribution):
  def __init__(self, *args, **kwargs):
    super(CustomRandomVariable, self).__init__(*args, **kwargs)

  def _log_prob(self, value):
    raise NotImplementedError("log_prob is not implemented")

  def _sample_n(self, n, seed=None):
    raise NotImplementedError("sample_n is not implemented")
\end{lstlisting}

One method that all Edward random variables call during instantiation is
\texttt{_sample_n()}.
It takes an integer \texttt{n} as input and outputs a tensor of shape
\texttt{(n,) + batch_shape + event_shape}.
To implement it, you can for example wrap a NumPy/SciPy function
inside the TensorFlow operation \texttt{tf.py_func()}.

For other methods and attributes one can implement, see the API documentation in
TensorFlow's
\texttt{Distribution} class.

\subsubsection{Advanced settings}

Sometimes the random variable you'd like to work with already exists
in Edward, but it is missing a particular feature. One hack is to
implement and overwrite the missing method. For example, to implement
your own sampling for \texttt{Poisson}:

\begin{lstlisting}[language=Python]
import edward as ed
from edward.models import Poisson
from scipy.stats import poisson

def _sample_n(self, n=1, seed=None):
  # define Python function which returns samples as a Numpy array
  def np_sample(lam, n):
    return poisson.rvs(mu=lam, size=n, random_state=seed).astype(np.float32)

  # wrap python function as tensorflow op
  val = tf.py_func(np_sample, [self.lam, n], [tf.float32])[0]
  # set shape from unknown shape
  batch_event_shape = self.get_batch_shape().concatenate(self.get_event_shape())
  shape = tf.concat([tf.expand_dims(n, 0),
                     tf.constant(batch_event_shape.as_list(), dtype=tf.int32)],
                     0)
  val = tf.reshape(val, shape)
  return val

Poisson._sample_n = _sample_n

sess = ed.get_session()
x = Poisson(lam=1.0)
sess.run(x.value())
## 1.0
sess.run(x.value())
## 4.0
\end{lstlisting}

(Note the function \texttt{np_sample} should broadcast correctly if
you'd like to work with non-scalar parameters; it is not correct in
this toy implementation.)

Sometimes the random variable you'd like to work with does not even
admit (easy) sampling, and you're only using it as a likelihood ``node'' rather
than as some prior to parameters of another random variable.
You can avoid having to implement \texttt{_sample_n} altogether:
after creating \texttt{CustomRandomVariable}, instantiate it with the
\texttt{value} argument:

\begin{lstlisting}[language=Python]
x = CustomRandomVariable(custom_params=params, value=tf.zeros_like(params))
\end{lstlisting}

This fixes the associated value of the random variable to a bunch of
zeros and avoids the \texttt{_sample_n} error that appears otherwise.
Make sure that the value matches the desired shape of the random
variable.

\subsection{Inference}

We describe how to perform inference in probabilistic models.

Suppose we have a model $p(\mathbf{x}, \mathbf{z}, \beta)$ of data $\mathbf{x}_{\text{train}}$ with latent variables $(\mathbf{z}, \beta)$.
Consider the posterior inference problem,
\begin{equation*}
q(\mathbf{z}, \beta)\approx p(\mathbf{z}, \beta\mid \mathbf{x}_{\text{train}}),
\end{equation*}
in which the task is to approximate the posterior
$p(\mathbf{z}, \beta\mid \mathbf{x}_{\text{train}})$
using a family of distributions, $q(\mathbf{z},\beta; \lambda)$,
indexed by parameters $\lambda$.

In Edward, let \texttt{z} and \texttt{beta} be latent variables in the model,
where we observe the random variable \texttt{x} with
data \texttt{x_train}.
Let \texttt{qz} and \texttt{qbeta} be random variables defined to
approximate the posterior.
We write this problem as follows:

\begin{lstlisting}[language=Python]
inference = ed.Inference({z: qz, beta: qbeta}, {x: x_train})
\end{lstlisting}

\texttt{Inference} is an abstract class which takes two inputs.  The
first is a collection of latent random variables \texttt{beta} and
\texttt{z}, along with ``posterior variables'' \texttt{qbeta} and
\texttt{qz}, which are associated to their respective latent
variables.  The second is a collection of observed random variables
\texttt{x}, which is associated to the data \texttt{x_train}.

Inference adjusts parameters of the distribution of \texttt{qbeta}
and \texttt{qz} to be close to the
posterior $p(\mathbf{z}, \beta\g \mbx_{\text{train}})$.

Running inference is as simple as running one method.

\begin{lstlisting}[language=Python]
inference = ed.Inference({z: qz, beta: qbeta}, {x: x_train})
inference.run()
\end{lstlisting}

Inference also supports fine control of the training procedure.

\begin{lstlisting}[language=Python]
inference = ed.Inference({z: qz, beta: qbeta}, {x: x_train})
inference.initialize()

tf.global_variables_initializer().run()

for _ in range(inference.n_iter):
  info_dict = inference.update()
  inference.print_progress(info_dict)

inference.finalize()
\end{lstlisting}

\texttt{initialize()} builds the algorithm's update rules
(computational graph) for $\lambda$; \\
\texttt{tf.global_variables_initializer().run()} initializes $\lambda$
(TensorFlow variables in the graph);
\texttt{update()} runs the graph once to update
$\lambda$, which is called in a loop until convergence;
\texttt{finalize()} runs any computation as the algorithm
terminates.

The \texttt{run()} method is a simple wrapper for this procedure.

\subsubsection{Other Settings}

We highlight other settings during inference.

\textbf{Model parameters}.
Model parameters are parameters in a model that we will always compute
point estimates for and not be uncertain about.
They are defined with \texttt{tf.Variable}s, where the inference
problem is
\begin{equation*}
\hat{\theta} \leftarrow^{\text{optimize}}
p(\mathbf{x}_{\text{train}}; \theta)
\end{equation*}

\begin{lstlisting}[language=Python]
from edward.models import Normal

theta = tf.Variable(0.0)
x = Normal(mu=tf.ones(10) * theta, sigma=1.0)

inference = ed.Inference({}, {x: x_train})
\end{lstlisting}

Only a subset of inference algorithms support estimation of model
parameters.
(Note also that this inference example does not have any latent
variables. It is only about estimating \texttt{theta} given that we
observe $\mathbf{x} = \mathbf{x}_{\text{train}}$. We can add them so
that inference is both posterior inference and parameter estimation.)

For example, model parameters are useful when applying neural networks
from high-level libraries such as Keras and TensorFlow Slim. See
the model compositionality subsection
for more details.

\textbf{Conditional inference}.
In conditional inference, only a subset of the posterior is inferred
while the rest are fixed using other inferences. The inference
problem is
\begin{equation*}
q(\beta)q(\mathbf{z})\approx
p(\mathbf{z}, \beta\mid\mathbf{x}_{\text{train}})
\end{equation*}
where parameters in $q(\beta)$ are estimated and $q(\mathbf{z})$ is
fixed.
In Edward, we enable conditioning by binding random variables to other
random variables in \texttt{data}.
\begin{lstlisting}[language=Python]
inference = ed.Inference({beta: qbeta}, {x: x_train, z: qz})
\end{lstlisting}

In the inference compositionality subsection,
we describe how to construct inference by composing
many conditional inference algorithms.

\textbf{Implicit prior samples}.
Latent variables can be defined in the model without any posterior
inference over them. They are implicitly marginalized out with a
single sample. The inference problem is
\begin{equation*}
q(\beta)\approx
p(\beta\mid\mathbf{x}_{\text{train}}, \mathbf{z}^*)
\end{equation*}
where $\mathbf{z}^*\sim p(\mathbf{z}\mid\beta)$ is a prior sample.

\begin{lstlisting}[language=Python]
inference = ed.Inference({beta: qbeta}, {x: x_train})
\end{lstlisting}

For example, implicit prior samples are useful for generative adversarial
networks. Their inference problem does not require any inference over
the latent variables; it uses samples from the prior.

\subsection*{Classes of Inference}

Inference is broadly classified under three classes: variational
inference, Monte Carlo, and exact inference.
We highlight how to use inference algorithms from each class.

As an example, we assume a mixture model with latent mixture
assignments \texttt{z}, latent cluster means \texttt{beta}, and
observations \texttt{x}:
\begin{equation*}
p(\mathbf{x}, \mathbf{z}, \beta)
=
\text{Normal}(\mathbf{x} \mid \beta_{\mathbf{z}}, \mathbf{I})
~
\text{Categorical}(\mathbf{z}\mid \pi)
~
\text{Normal}(\beta\mid \mathbf{0}, \mathbf{I}).
\end{equation*}

\subsubsection{Variational Inference}

In variational inference, the idea is to posit a family of approximating
distributions and to find the closest member in the family to the
posterior \citep{jordan1999introduction}.
We write an approximating family,
\begin{align*}
q(\beta;\mu,\sigma) &= \text{Normal}(\beta; \mu,\sigma), \\[1.5ex]
q(\mathbf{z};\pi) &= \text{Categorical}(\mathbf{z};\pi),
\end{align*}
using TensorFlow variables to represent its parameters
$\lambda=\{\pi,\mu,\sigma\}$.
\begin{lstlisting}[language=Python]
from edward.models import Categorical, Normal

qbeta = Normal(mu=tf.Variable(tf.zeros([K, D])),
               sigma=tf.exp(tf.Variable(tf.zeros[K, D])))
qz = Categorical(logits=tf.Variable(tf.zeros[N, K]))

inference = ed.VariationalInference({beta: qbeta, z: qz}, data={x: x_train})
\end{lstlisting}
Given an objective function, variational inference optimizes the
family with respect to \texttt{tf.Variable}s.

Specific variational inference algorithms inherit from
the \texttt{VariationalInference} class to define their own methods, such as a
loss function and gradient.
For example, we represent
MAP
estimation with an approximating family (\texttt{qbeta} and
\texttt{qz}) of \texttt{PointMass} random variables, i.e., with all
probability mass concentrated at a point.
\begin{lstlisting}[language=Python]
from edward.models import PointMass

qbeta = PointMass(params=tf.Variable(tf.zeros([K, D])))
qz = PointMass(params=tf.Variable(tf.zeros(N)))

inference = ed.MAP({beta: qbeta, z: qz}, data={x: x_train})
\end{lstlisting}
\texttt{MAP} inherits from \texttt{VariationalInference} and defines a
loss function and update rules; it uses existing optimizers inside
TensorFlow.

\subsubsection{Monte Carlo}

Monte Carlo approximates the posterior using samples
\citep{robert1999monte}. Monte Carlo is an inference where the
approximating family is an empirical distribution,
\begin{align*}
q(\beta; \{\beta^{(t)}\})
&= \frac{1}{T}\sum_{t=1}^T \delta(\beta, \beta^{(t)}), \\[1.5ex]
q(\mathbf{z}; \{\mathbf{z}^{(t)}\})
&= \frac{1}{T}\sum_{t=1}^T \delta(\mathbf{z}, \mathbf{z}^{(t)}).
\end{align*}
The parameters are $\lambda=\{\beta^{(t)},\mathbf{z}^{(t)}\}$.
\begin{lstlisting}[language=Python]
from edward.models import Empirical

T = 10000  # number of samples
qbeta = Empirical(params=tf.Variable(tf.zeros([T, K, D]))
qz = Empirical(params=tf.Variable(tf.zeros([T, N]))

inference = ed.MonteCarlo({beta: qbeta, z: qz}, data={x: x_train})
\end{lstlisting}
Monte Carlo
algorithms proceed by updating one sample $\beta^{(t)},\mathbf{z}^{(t)}$ at a time in the empirical
approximation.
Monte Carlo algorithms proceed by updating one sample
$\beta^{(t)},\mbz^{(t)}$ at a time in the empirical approximation.
Markov chain Monte Carlo does this sequentially to update
the current sample (index $t$ of \texttt{tf.Variable}s) conditional on
the last sample (index $t-1$ of \texttt{tf.Variable}s).
Specific Monte Carlo samplers determine the update rules;
they can use gradients such as in Hamiltonian Monte Carlo
\citep{neal2011mcmc} and graph
structure such as in sequential Monte Carlo \citep{doucet2001introduction}.

\subsubsection{Exact Inference}

This approach also extends to algorithms that usually require tedious
algebraic manipulation.  With symbolic algebra on the nodes of the
computational graph, we can uncover conjugacy relationships between
random variables.  Users can then integrate out variables to
automatically derive classical Gibbs \citep{gelfand1990sampling},
mean-field updates \citep{bishop2006pattern}, and exact inference.

\subsection*{Composing Inferences}

Core to Edward's design is compositionality. Compositionality enables
fine control of inference, where we can write inference as a
collection of separate inference programs.

We outline how to write popular classes of compositional inferences
using Edward: hybrid algorithms and message passing algorithms.
We use the running example of a mixture model
with latent mixture assignments \texttt{z}, latent cluster means
\texttt{beta}, and observations \texttt{x}.

\subsubsection{Hybrid algorithms}

Hybrid algorithms leverage different inferences for each latent
variable in the posterior.
As an example, we demonstrate variational EM, with an approximate
E-step over local variables and an M-step over global variables.
We alternate with one update of each \citep{neal1993new}.

\begin{lstlisting}[language=Python]
from edward.models import Categorical, PointMass

qbeta = PointMass(params=tf.Variable(tf.zeros([K, D])))
qz = Categorical(logits=tf.Variable(tf.zeros[N, K]))

inference_e = ed.VariationalInference({z: qz}, data={x: x_data, beta: qbeta})
inference_m = ed.MAP({beta: qbeta}, data={x: x_data, z: qz})
...
for _ in range(10000):
  inference_e.update()
  inference_m.update()
\end{lstlisting}

In \texttt{data}, we include bindings of prior latent variables
(\texttt{z} or \texttt{beta}) to posterior latent variables
(\texttt{qz} or \texttt{qbeta}). This performs conditional inference,
where only a subset of the posterior is inferred while the rest are
fixed using other inferences.

This extends to many algorithms: for example,
exact EM for exponential families;
contrastive divergence \citep{hinton2002training};
pseudo-marginal and ABC methods \citep{andrieu2009pseudo};
Gibbs sampling within variational inference \citep{wang2012truncation};
Laplace variational inference \citep{wang2013variational};
and
structured variational auto-encoders \citep{johnson2016composing}.

\subsubsection{Message passing algorithms}

Message passing algorithms operate on the posterior distribution using
a collection of local inferences \citep{koller2009probabilistic}.
As an example, we demonstrate expectation propagation. We split a
mixture model to be over two random variables \texttt{x1} and
\texttt{x2} along with their latent mixture assignments \texttt{z1}
and \texttt{z2}.

\begin{lstlisting}[language=Python]
from edward.models import Categorical, Normal

N1 = 1000  # number of data points in first data set
N2 = 2000  # number of data points in second data set
D = 2  # data dimension
K = 5  # number of clusters

# MODEL
beta = Normal(mu=tf.zeros([K, D]), sigma=tf.ones([K, D]))
z1 = Categorical(logits=tf.zeros([N1, K]))
z2 = Categorical(logits=tf.zeros([N2, K]))
x1 = Normal(mu=tf.gather(beta, z1), sigma=tf.ones([N1, D]))
x2 = Normal(mu=tf.gather(beta, z2), sigma=tf.ones([N2, D]))

# INFERENCE
qbeta = Normal(mu=tf.Variable(tf.zeros([K, D])),
               sigma=tf.nn.softplus(tf.Variable(tf.zeros([K, D]))))
qz1 = Categorical(logits=tf.Variable(tf.zeros[N1, K]))
qz2 = Categorical(logits=tf.Variable(tf.zeros[N2, K]))

inference_z1 = ed.KLpq({beta: qbeta, z1: qz1}, {x1: x1_train})
inference_z2 = ed.KLpq({beta: qbeta, z2: qz2}, {x2: x2_train})
...
for _ in range(10000):
  inference_z1.update()
  inference_z2.update()
\end{lstlisting}

We alternate updates for each local inference, where the global
posterior factor $q(\beta)$ is shared across both inferences
\citep{gelman2014expectation}.

With TensorFlow's distributed training, compositionality
enables \emph{distributed} message passing over a cluster with many
workers. The computation can be further sped up with the use of GPUs
via data and model parallelism.

This extends to many algorithms: for example,
classical message passing, which performs exact local inferences;
Gibbs sampling, which draws samples from conditionally conjugate
inferences \citep{geman1984stochastic};
expectation propagation, which locally minimizes
$\text{KL}(p || q)$ over exponential families \citep{minka2001expectation};
integrated nested Laplace
approximation, which performs local Laplace approximations
\citep{rue2009approximate};
and
all the instantiations of EP-like algorithms in
\citet{gelman2014expectation}.

In the above, we perform local inferences split over individual random
variables. At the moment, Edward does not support local inferences
within a random variable itself. We cannot do local inferences when
representing the random variable for all data points and their cluster
membership as \texttt{x} and \texttt{z} rather than \texttt{x1},
\texttt{x2}, \texttt{z1}, and \texttt{z2}.

\subsection*{Data Subsampling}

Running algorithms which require the full data set for each update
can be expensive when the data is large. In order to scale inferences,
we can do \emph{data subsampling}, i.e., update inference using
only a subsample of data at a time.
(Note that only certain algorithms can support data subsampling such as
\texttt{MAP}, \texttt{KLqp}, and \texttt{SGLD}.)

\subsubsection{Subgraphs}

\begin{figure}[!htb]
\centering
\begin{tikzpicture}

  \node[latent]               (beta)      {$\beta$} ;
  \node[left=0.5cm of beta]               (betaph)      {} ;
  \node[latent, below=1.0cm of betaph] (z)    {$z_m$} ;
  \node[obs, right=1.0cm of z]      (x)        {$x_m$} ;

  \edge{beta}{x};
  \edge{z}{x};

  \plate[inner sep=0.3cm,
    label={[xshift=-15pt,yshift=15pt]south east:$M$}] {plate1} {
    (x)(z)
  } {};

\end{tikzpicture}
\caption{Data subsampling with a hierarchical model. We define
a subgraph of the full model, forming a plate of size $M$ rather than
$N$.}
\end{figure}
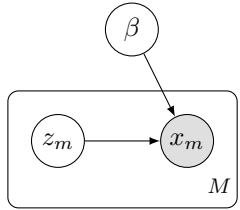

In the subgraph setting, we do data subsampling while working with a
subgraph of the full model. This setting is necessary when the data
and model do not fit in memory.
It is scalable in that both the
algorithm's computational complexity (per iteration) and memory
complexity are independent of the data set size.

For example, consider a hierarchical model,
\begin{equation*}
p(\mathbf{x}, \mathbf{z}, \beta)
= p(\beta) \prod_{n=1}^N p(z_n \mid \beta) p(x_n \mid z_n, \beta),
\end{equation*}
where there are latent variables $z_n$ for
each data point $x_n$ (local variables) and latent variables $\beta$
which are shared across data points (global variables).

To avoid memory issues, we work on only a subgraph of the model,
\begin{equation*}
p(\mathbf{x}, \mathbf{z}, \beta)
= p(\beta) \prod_{m=1}^M p(z_m \mid \beta) p(x_m \mid z_m, \beta).
\end{equation*}
More concretely, we define a mixture of Gaussians over
$D$-dimensional data $\{x_n\}\in\mathbb{R}^{N\times D}$. There are $K$
latent cluster means $\{\beta_k\}\in\mathbb{R}^{K\times D}$ and a
membership assignment $z_n\in\{0,\ldots,K-1\}$ for each data point
$x_n$.

\begin{lstlisting}[language=Python]
N = 10000000  # data set size
M = 128  # minibatch size
D = 2  # data dimensionality
K = 5  # number of clusters

beta = Normal(mu=tf.zeros([K, D]), sigma=tf.ones([K, D]))
z = Categorical(logits=tf.zeros([M, K]))
x = Normal(mu=tf.gather(beta, z), sigma=tf.ones([M, D]))
\end{lstlisting}

For inference, the variational model is
\begin{equation*}
q(\mathbf{z}, \beta) =
q(\beta; \lambda) \prod_{n=1}^N q(z_n \mid \beta; \gamma_n),
\end{equation*}
parameterized by $\{\lambda, \{\gamma_n\}\}$.
Again, we work on only a subgraph of the model,
\begin{equation*}
q(\mathbf{z}, \beta) =
q(\beta; \lambda) \prod_{m=1}^M q(z_m \mid \beta; \gamma_m).
\end{equation*}
parameterized by $\{\lambda, \{\gamma_m\}\}$. Importantly, only $M$
parameters are stored in memory for $\{\gamma_m\}$ rather than $N$.

\begin{lstlisting}[language=Python]
qbeta = Normal(mu=tf.Variable(tf.zeros([K, D])),
               sigma=tf.nn.softplus(tf.Variable(tf.zeros[K, D])))
qz_variables = tf.Variable(tf.zeros([M, K]))
qz = Categorical(logits=qz_variables)
\end{lstlisting}

We will perform inference with \texttt{KLqp}, a variational method
that minimizes the divergence measure $\text{KL}(q\| p)$.

We instantiate two algorithms: a global inference over $\beta$ given
the subset of $\mathbf{z}$ and a local inference over the subset of
$\mathbf{z}$ given $\beta$.
We also pass in a TensorFlow placeholder \texttt{x_ph} for the data,
so we can change the data at each step. (Alternatively,
{batch tensors} can be used.)

\begin{lstlisting}[language=Python]
x_ph = tf.placeholder(tf.float32, [M])
inference_global = ed.KLqp({beta: qbeta}, data={x: x_ph, z: qz})
inference_local = ed.KLqp({z: qz}, data={x: x_ph, beta: qbeta})
\end{lstlisting}

We initialize the algorithms with the \texttt{scale} argument, so that
computation on \texttt{z} and \texttt{x} will be scaled appropriately.
This enables unbiased estimates for stochastic gradients.

\begin{lstlisting}[language=Python]
inference_global.initialize(scale={x: float(N) / M, z: float(N) / M})
inference_local.initialize(scale={x: float(N) / M, z: float(N) / M})
\end{lstlisting}

Conceptually, the scale argument represents scaling for each random
variable’s plate, as if we had seen that random variable $N/M$ as many
times.

We now run inference, assuming there is a \texttt{next_batch} function
which provides the next batch of data.

\begin{lstlisting}[language=Python]
qz_init = tf.initialize_variables([qz_variables])
for _ in range(1000):
  x_batch = next_batch(size=M)
  for _ in range(10):  # make local inferences
    inference_local.update(feed_dict={x_ph: x_batch})

  # update global parameters
  inference_global.update(feed_dict={x_ph: x_batch})
  # reinitialize the local factors
  qz_init.run()
\end{lstlisting}

After each iteration, we also reinitialize the parameters for
$q(\mathbf{z}\mid\beta)$; this is because we do inference on a new
set of local variational factors for each batch.

This demo readily applies to other inference algorithms such as
\texttt{SGLD} (stochastic gradient Langevin dynamics): simply
replace \texttt{qbeta} and \texttt{qz} with \texttt{Empirical} random
variables; then call \texttt{ed.SGLD} instead of \texttt{ed.KLqp}.

\subsubsection{Advanced settings}

If the parameters fit in memory, one can avoid having to reinitialize
local parameters or read/write from disk.  To do this, define the full
set of parameters and index them into the local posterior factors.

\begin{lstlisting}[language=Python]
qz_variables = tf.Variable(tf.zeros([N, K]))
idx_ph = tf.placeholder(tf.int32, [M])
qz = Categorical(logits=tf.gather(qz_variables, idx_ph))
\end{lstlisting}

We define an index placeholder \texttt{idx_ph}. It will be fed index
values at runtime to determine which parameters correspond to a given
data subsample.

An alternative approach to reduce memory complexity is to use an
inference network \citep{dayan1995helmholtz}, also known as
amortized inference \citep{stuhlmuller2013learning}.  This can be
applied using a global parameterization of $q(\mathbf{z}, \beta)$.

In streaming data, or online inference, the size of the data $N$
may be unknown, or conceptually the size of the data may be
infinite and at any time in which we query parameters from the online
algorithm, the outputted parameters are from having processed as many
data points up to that time.
The approach of Bayesian filtering
\citep{doucet2000on,broderick2013streaming} can be applied in Edward using
recursive posterior inferences; the approach of population posteriors
\citep{mcinerney2015population} is readily applicable from the subgraph
setting.

In other settings, working on a subgraph of the model does not
apply, such as in time series models when we want to
preserve dependencies across time steps in our variational model.
Approaches in the literature can be applied in Edward
\citep{binder1997space,johnson2014stochastic,foti2014stochastic}.

\subsubsection*{Development of Inference Methods}

Edward uses class inheritance to provide a hierarchy of inference
methods. This enables fast experimentation on top of existing
algorithms, whether it be developing new black box algorithms or
new model-specific algorithms.

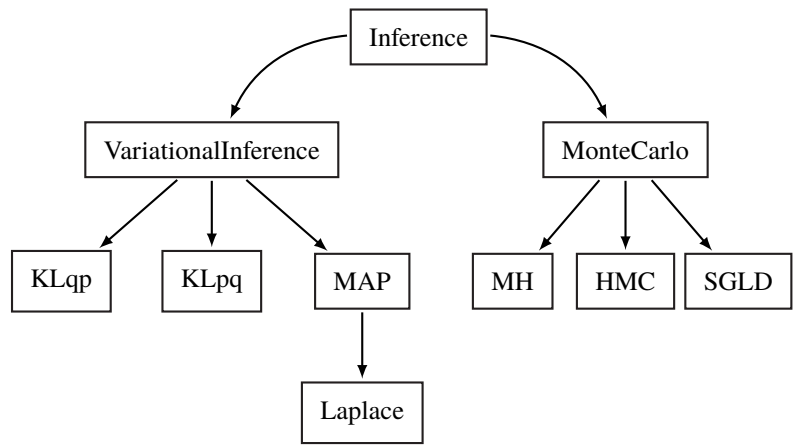
\begin{figure}[!htb]
\centering
\begin{tikzpicture}
    \begin{pgfonlayer}{nodelayer}
        \node [style=box] (0) at (2.25, 2) {Inference};
        \node [style=box] (1) at (-0.5, 0.5) {VariationalInference};
        \node [style=box] (2) at (5, 0.5) {MonteCarlo};
        \node [style=box] (3) at (-2.5, -1.25) {KLqp};
        \node [style=box] (4) at (-0.5, -1.25) {KLpq};
        \node [style=box] (5) at (1.5, -1.25) {MAP};
        \node [style=box] (6) at (1.5, -3) {Laplace};
        \node [style=box] (7) at (3.5, -1.25) {MH};
        \node [style=box] (8) at (5, -1.25) {HMC};
        \node [style=box] (9) at (6.5, -1.25) {SGLD};
    \end{pgfonlayer}
    \begin{pgfonlayer}{edgelayer}
        \draw [style=arrow, bend right] (0) to (1);
        \draw [style=arrow] (1) to (3);
        \draw [style=arrow] (1) to (4);
        \draw [style=arrow] (1) to (5);
        \draw [style=arrow] (5) to (6);
        \draw [style=arrow, bend left] (0) to (2);
        \draw [style=arrow] (2) to (7);
        \draw [style=arrow] (2) to (8);
        \draw [style=arrow] (2) to (9);
    \end{pgfonlayer}
\end{tikzpicture}
\caption{Dependency graph of inference methods. Nodes are classes in Edward
and arrows represent class inheritance.}
\end{figure}

There is a base class \texttt{Inference}, from which all inference
methods are derived from. Note that \texttt{Inference} says nothing
about the class of models that an algorithm must work with. One can
build inference algorithms which are tailored to a restricted class of
models available in Edward (such as differentiable models or
conditionally conjugate models), or even tailor it to a single model.
The algorithm can raise an error if the model is outside this class.

We organize inference under two paradigms:
\texttt{VariationalInference} and \texttt{MonteCarlo} (or more plainly,
optimization and sampling). These inherit from \texttt{Inference} and each
have their own default methods.

For example, developing a new variational inference algorithm is as simple as
inheriting from \texttt{VariationalInference} and writing a
\texttt{build_loss_and_gradients()} method. \texttt{VariationalInference} implements many default methods such
as \texttt{initialize()} with options for an optimizer.
For example, see the
importance weighted variational inference
script.%
\footnote{\url{https://github.com/blei-lab/edward/blob/master/examples/iwvi.py}}

\subsection{Criticism}

We can never validate whether a model is true. In practice, ``all
models are wrong'' \citep{box1976science}. However, we can try to
uncover where the model goes wrong. Model criticism helps justify the
model as an approximation or point to good directions for revising the
model.

Edward explores model criticism using
\begin{itemize}
  \item point-based evaluations, such as mean squared error or
  classification accuracy;
  \item posterior predictive checks, for making probabilistic
  assessments of the model fit using discrepancy functions.
\end{itemize}

We describe them in detail below.

\subsubsection{Point-based evaluations}

A point-based evaluation is a scalar-valued metric for assessing
trained models \citep{winkler1996scoring,gneiting2007strictly}.
For example, we can assess models for classification
by predicting the label for each observation in the data and comparing
it to their true labels. Edward implements a variety of metrics, such
as classification error and mean absolute error.

Formally, point prediction in probabilistic models is given by
taking the mean of the posterior predictive distribution,
\begin{align*}
  p(\mathbf{x}_\text{new} \mid \mathbf{x})
  &=
  \int
  p(\mathbf{x}_\text{new} \mid \mathbf{z})
  p(\mathbf{z} \mid \mathbf{x})
  \text{d} \mathbf{z}.
\end{align*}
The model's posterior predictive can be used to generate new data
given past observations and can also make predictions on new data
given past observations.
It is formed by calculating the likelihood of the new data, averaged
over every set of latent variables according to the posterior
distribution.

\subsubsection{Implementation}

To evaluate inferred models, we first form the posterior
predictive distribution. A helpful utility function for this is
\texttt{copy}. For example,
assume the model defines a likelihood \texttt{x} connected to a prior
\texttt{z}. The posterior predictive distribution is
\begin{lstlisting}[language=Python]
x_post = ed.copy(x, {z: qz})
\end{lstlisting}
Here, we copy the likelihood node \texttt{x} in the graph and replace dependence
on the prior \texttt{z} with dependence on the inferred posterior \texttt{qz}.

The \texttt{ed.evaluate()} method takes as input a set of metrics to
evaluate, and a data dictionary. As with inference, the data dictionary binds the
observed random variables in the model to realizations: in this case,
it is the posterior predictive random variable of outputs \texttt{y_post} to
\texttt{y_train} and a placeholder for inputs \texttt{x} to
\texttt{x_train}.
\begin{lstlisting}[language=Python]
ed.evaluate('categorical_accuracy', data={y_post: y_train, x: x_train})
ed.evaluate('mean_absolute_error', data={y_post: y_train, x: x_train})
\end{lstlisting}
The \texttt{data} can be data held-out from training time, making it
easy to implement cross-validated techniques.

Point-based evaluation applies generally to any setting, including
unsupervised tasks. For example, we can evaluate the likelihood of
observing the data.
\begin{lstlisting}[language=Python]
ed.evaluate('log_likelihood', data={x_post: x_train})
\end{lstlisting}

It is common practice to criticize models with data held-out from
training. To do this, we first perform inference over any local latent
variables of the held-out data, fixing the global variables.  Then we
make predictions on the held-out data.

\begin{lstlisting}[language=Python]
from edward.models import Categorical

# create local posterior factors for test data, assuming test data
# has N_test many data points
qz_test = Categorical(logits=tf.Variable(tf.zeros[N_test, K]))

# run local inference conditional on global factors
inference_test = ed.Inference({z: qz_test}, data={x: x_test, beta: qbeta})
inference_test.run()

# build posterior predictive on test data
x_post = ed.copy(x, {z: qz_test, beta: qbeta}})
ed.evaluate('log_likelihood', data={x_post: x_test})
\end{lstlisting}

Point-based evaluations are formally known as scoring rules
in decision theory. Scoring rules are useful for model comparison, model
selection, and model averaging.

\subsubsection{Posterior predictive checks}

Posterior predictive checks (PPCs)
analyze the degree to which data generated from the model deviate from
data generated from the true distribution. They can be used either
numerically to quantify this degree, or graphically to visualize this
degree. PPCs can be thought of as a probabilistic generalization of
point-based evaluations
\citep{box1980sampling,rubin1984bayesianly,meng1994posterior,gelman1996posterior}.

PPCs focus on the posterior predictive distribution
\begin{align*}
  p(\mathbf{x}_\text{new} \mid \mathbf{x})
  &=
  \int
  p(\mathbf{x}_\text{new} \mid \mathbf{z})
  p(\mathbf{z} \mid \mathbf{x})
  \text{d} \mathbf{z}.
\end{align*}
%
The simplest PPC works by applying a test statistic on new data
generated from the posterior predictive, such as
$T(\mathbf{x}_\text{new}) = \max(\mathbf{x}_\text{new})$.  Applying
$T(\mathbf{x}_\text{new})$ to new data over many data replications
induces a distribution. We compare this distribution to the test
statistic applied to the real data $T(\mathbf{x})$.

In the figure, $T(\mathbf{x})$ falls in a low probability region of
this reference distribution. This indicates that the model fits the
data poorly according to this check; this suggests an area of
improvement for the model.

More generally, the test statistic can also be a function of the
model's latent variables $T(\mathbf{x}, \mathbf{z})$, known as a
discrepancy function.  Examples of discrepancy functions are the
metrics used for point-based evaluation. We can now interpret the
point-based evaluation as a special case of PPCs: it simply calculates
$T(\mathbf{x}, \mathbf{z})$ over the real data and without a reference
distribution in mind. A reference distribution allows us to make
probabilistic statements about the point, in reference to an overall
distribution.

\subsubsection{Implementation}

To evaluate inferred models, we first form the posterior
predictive distribution. A helpful utility function for this is
\texttt{copy}. For example,
assume the model defines a likelihood \texttt{x} connected to a prior
\texttt{z}. The posterior predictive distribution is
\begin{lstlisting}[language=Python]
x_post = ed.copy(x, {z: qz})
\end{lstlisting}
Here, we copy the likelihood node \texttt{x} in the graph and replace dependence
on the prior \texttt{z} with dependence on the inferred posterior \texttt{qz}.

The \texttt{ed.ppc()} method provides a scaffold for studying
various discrepancy functions.
\begin{lstlisting}[language=Python]
def T(xs, zs):
  return tf.reduce_mean(xs[x_post])

ed.ppc(T, data={x_post: x_train})
\end{lstlisting}
The discrepancy can also take latent variables as input, which we pass
into the PPC.
\begin{lstlisting}[language=Python]
def T(xs, zs):
  return tf.reduce_mean(tf.cast(zs['z'], tf.float32))

ppc(T, data={y_post: y_train, x_ph: x_train},
    latent_vars={'z': qz, 'beta': qbeta})
\end{lstlisting}


PPCs are an excellent tool for revising models, simplifying or
expanding the current model as one examines how well it fits the data.
They are inspired by prior checks and classical hypothesis
testing, under the philosophy that models should be
criticized under the frequentist perspective of large sample
assessment.

PPCs can also be applied to tasks such as hypothesis testing, model
comparison, model selection, and model averaging.  It's important to
note that while they can be applied as a form of Bayesian hypothesis
testing, hypothesis testing is generally not recommended: binary
decision making from a single test is not as common a use case as one
might believe. We recommend performing many PPCs to get a holistic
understanding of the model fit.

\clearpage
\section{End-to-end Examples}
\label{sec:examples}

\subsection{Bayesian Linear Regression}

In supervised learning, the task is to infer hidden structure from
labeled data, comprised of training examples $\{(x_n, y_n)\}$.
Regression (typically) means the output $y$ takes continuous values.

\subsubsection{Data}

Simulate training and test sets of $500$ data points. They comprise
pairs of inputs $\mathbf{x}_n\in\mathbb{R}^{5}$ and outputs
$y_n\in\mathbb{R}$. They have a linear dependence of
\begin{align*}
  \mathbf{w}_{\text{true}}
  &=
  (-1.25, 4.51, 2.32, 0.99, -3.44).
\end{align*}
with normally distributed noise.

\begin{lstlisting}
def build_toy_dataset(N, w, noise_std=0.1):
  D = len(w)
  x = np.random.randn(N, D).astype(np.float32)
  y = np.dot(x, w) + np.random.normal(0, noise_std, size=N)
  return x, y

N = 500  # number of  data points
D = 5  # number of  features

w_true = 10 * (np.random.rand(D) - 0.5)
X_train, y_train = build_toy_dataset(N, w_true)
X_test, y_test = build_toy_dataset(N, w_true)
\end{lstlisting}

\subsubsection{Model}

Posit the model as Bayesian linear regression. It relates
outputs $y\in\mathbb{R}$, also known as the response, given
a vector of inputs
$\mathbf{x}\in\mathbb{R}^D$, also known as the features or covariates.
The model assumes a
linear relationship between these two random variables
\citep{murphy2012machine}.

For a set of $N$ data points $(\mathbf{X},\mathbf{y})=\{(\mathbf{x}_n, y_n)\}$,
the model posits the following conditional relationships:
\begin{align*}
  p(\mathbf{w})
  &=
  \text{Normal}(\mathbf{w} \mid \mathbf{0}, \sigma_w^2\mathbf{I}),
  \\[1.5ex]
  p(b)
  &=
  \text{Normal}(b \mid 0, \sigma_b^2),
  \\
  p(\mathbf{y} \mid \mathbf{w}, b, \mathbf{X})
  &=
  \prod_{n=1}^N
  \text{Normal}(y_n \mid \mathbf{x}_n^\top\mathbf{w} + b, \sigma_y^2).
\end{align*}
The latent variables are the linear model's weights $\mathbf{w}$ and
intercept $b$, also known as the bias.
Assume $\sigma_w^2,\sigma_b^2$ are known prior variances and $\sigma_y^2$ is a
known likelihood variance. The mean of the likelihood is given by a
linear transformation of the inputs $\mathbf{x}_n$.

\begin{lstlisting}
X = tf.placeholder(tf.float32, [N, D])
w = Normal(mu=tf.zeros(D), sigma=tf.ones(D))
b = Normal(mu=tf.zeros(1), sigma=tf.ones(1))
y = Normal(mu=ed.dot(X, w) + b, sigma=tf.ones(N))
\end{lstlisting}

\subsubsection{Inference}

We now turn to inferring the posterior using variational inference.
Define the variational model to be a fully factorized normal across
the weights.
\begin{lstlisting}
qw = Normal(mu=tf.Variable(tf.random_normal([D])),
            sigma=tf.nn.softplus(tf.Variable(tf.random_normal([D]))))
qb = Normal(mu=tf.Variable(tf.random_normal([1])),
            sigma=tf.nn.softplus(tf.Variable(tf.random_normal([1]))))
\end{lstlisting}

\begin{figure}[htb]
\centering
\input{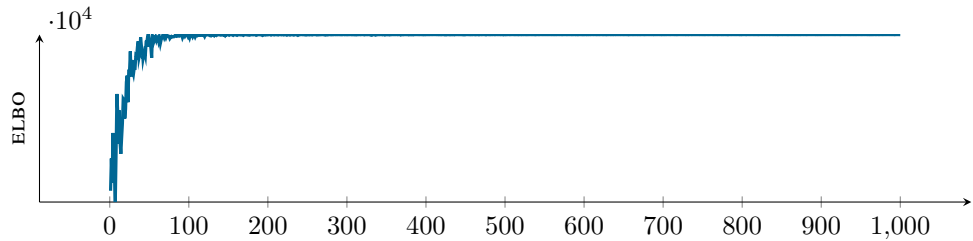}
\caption{The evidence lower bound (\textsc{elbo}) as a function of iterations.
Variational inference maximizes this quantity iteratively; in this case, the
algorithm appears to have converged in approximately 200 iterations.}
\label{fig:supervised-elbo}
\end{figure}

Run variational inference with the Kullback-Leibler divergence, using a
default of $1000$ iterations.
\begin{lstlisting}
inference = ed.KLqp({w: qw, b: qb}, data={X: X_train, y: y_train})
inference.run()
\end{lstlisting}
In this case \texttt{KLqp} defaults to minimizing the
$\text{KL}(q\|p)$ divergence measure using the reparameterization
gradient.
Minimizing this divergence metric is equivalent to maximizing the
evidence lower bound (\textsc{elbo}). Figure\nobreakspace \ref{fig:supervised-elbo} shows the
progression of the \textsc{elbo} across iterations; variational inference
appears to converge in approximately 200 iterations.

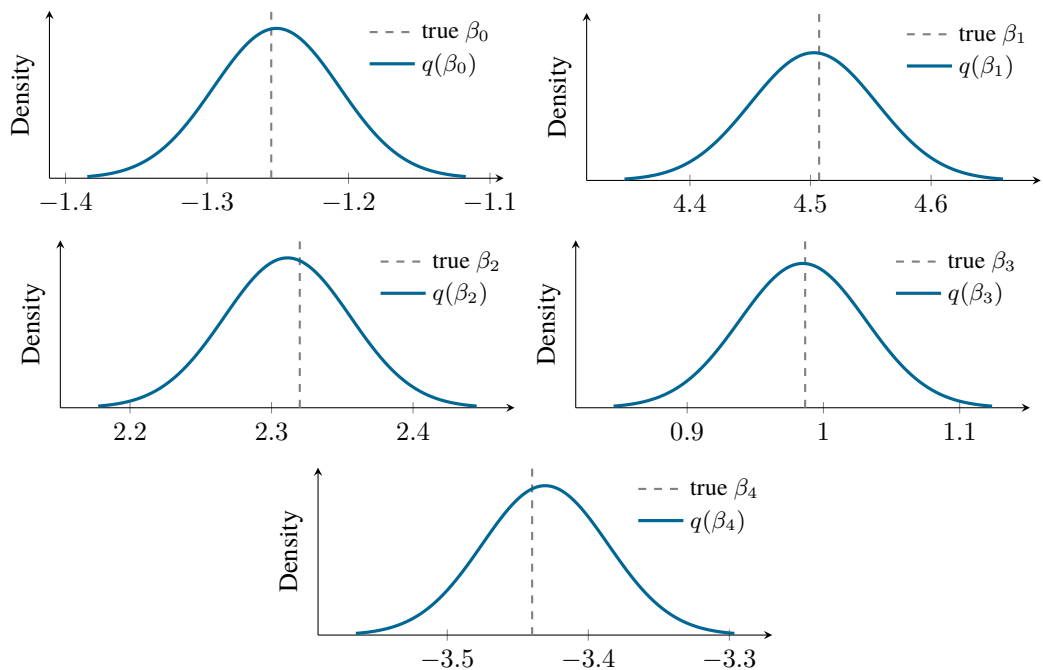
\begin{figure}[htb]
\centering
\begin{tikzpicture}

\begin{axis}[
width=3.0in,
height=1.5in,
axis lines=left,
enlarge x limits=0.1,
ylabel={Density},
ylabel near ticks,
ytick=\empty,
legend entries={{true $\beta_0$}, {$q(\beta_0)$}},
legend style={draw=none, at={(1.0,1.0)}, anchor=north east,font=\small},
legend cell align=left,
]
\addplot [thick, dashed, gray]
table {%
-1.25459881152638 0
-1.25459881152638 10
};
\addplot [very thick, MidnightBlue]
table {%
-1.3846001252532 0.0994103432712143
-1.38189822457957 0.119013603277019
-1.37919632390593 0.141960143663186
-1.3764944232323 0.168710087816997
-1.37379252255866 0.199765491467325
-1.37109062188503 0.23567020520197
-1.36838872121139 0.277008877312448
-1.36568682053776 0.324404962879766
-1.36298491986412 0.378517608358809
-1.36028301919048 0.440037289017542
-1.35758111851685 0.509680089993057
-1.35487921784321 0.588180540878532
-1.35217731716958 0.676282938940386
-1.34947541649594 0.774731127379731
-1.34677351582231 0.884256732367532
-1.34407161514867 1.00556590551315
-1.34136971447504 1.1393246663063
-1.3386678138014 1.28614299094489
-1.33596591312777 1.44655784857285
-1.33326401245413 1.62101544176614
-1.3305621117805 1.80985296332854
-1.32786021110686 2.01328023408348
-1.32515831043323 2.23136163420829
-1.32245640975959 2.46399878150339
-1.31975450908596 2.71091444157125
-1.31705260841232 2.97163817504537
-1.31435070773869 3.245494233803
-1.31164880706505 3.53159220986514
-1.30894690639142 3.82882091618547
-1.30624500571778 4.13584593700778
-1.30354310504415 4.45111122675914
-1.30084120437051 4.77284506101201
-1.29813930369688 5.0990705520425
-1.29543740302324 5.42762083676813
-1.29273550234961 5.75615892887182
-1.29003360167597 6.08220210282844
-1.28733170100234 6.40315054899689
-1.2846298003287 6.71631990999255
-1.28192789965507 7.01897718355629
-1.27922599898143 7.30837936054445
-1.2765240983078 7.58181406287413
-1.27382219763416 7.83664135941491
-1.27112029696053 8.07033587163715
-1.26841839628689 8.28052823841812
-1.26571649561326 8.46504499312375
-1.26301459493962 8.62194591739624
-1.26031269426599 8.74955797547933
-1.25761079359235 8.84650499987558
-1.25490889291872 8.91173239208829
-1.25220699224508 8.94452621859864
-1.24950509157145 8.94452621859864
-1.24680319089781 8.91173239208829
-1.24410129022418 8.84650499987558
-1.24139938955054 8.74955797547933
-1.23869748887691 8.62194591739624
-1.23599558820327 8.46504499312375
-1.23329368752964 8.28052823841812
-1.230591786856 8.07033587163715
-1.22788988618237 7.83664135941491
-1.22518798550873 7.58181406287413
-1.2224860848351 7.30837936054445
-1.21978418416146 7.01897718355629
-1.21708228348783 6.71631990999255
-1.21438038281419 6.40315054899689
-1.21167848214056 6.08220210282844
-1.20897658146692 5.75615892887182
-1.20627468079329 5.42762083676813
-1.20357278011965 5.0990705520425
-1.20087087944602 4.77284506101201
-1.19816897877238 4.45111122675914
-1.19546707809875 4.13584593700778
-1.19276517742511 3.82882091618547
-1.19006327675147 3.53159220986514
-1.18736137607784 3.245494233803
-1.1846594754042 2.97163817504537
-1.18195757473057 2.71091444157125
-1.17925567405693 2.46399878150339
-1.1765537733833 2.23136163420829
-1.17385187270966 2.01328023408348
-1.17114997203603 1.80985296332852
-1.16844807136239 1.62101544176614
-1.16574617068876 1.44655784857284
-1.16304427001512 1.28614299094489
-1.16034236934149 1.1393246663063
-1.15764046866785 1.00556590551315
-1.15493856799422 0.884256732367532
-1.15223666732058 0.774731127379731
-1.14953476664695 0.676282938940386
-1.14683286597331 0.588180540878532
-1.14413096529968 0.509680089993057
-1.14142906462604 0.440037289017542
-1.13872716395241 0.378517608358809
-1.13602526327877 0.324404962879766
-1.13332336260514 0.277008877312448
-1.1306214619315 0.23567020520197
-1.12791956125787 0.199765491467325
-1.12521766058423 0.168710087816997
-1.1225157599106 0.141960143663186
-1.11981385923696 0.119013603277019
-1.11711195856333 0.0994103432712143
};
\end{axis}

\end{tikzpicture}
\begin{tikzpicture}

\begin{axis}[
width=3.0in,
height=1.5in,
axis lines=left,
enlarge x limits=0.1,
ylabel={Density},
ylabel near ticks,
ytick=\empty,
legend entries={{true $\beta_1$}, {$q(\beta_1)$}},
legend style={draw=none, at={(1.0,1.0)}, anchor=north east,font=\small},
legend cell align=left,
]
\addplot [thick, dashed, gray]
table {%
4.50714306409916 0
4.50714306409916 10
};
\addplot [very thick, MidnightBlue]
table {%
4.34573828801513 0.0847196368493868
4.34890870851549 0.10142595748068
4.35207912901586 0.120981493700508
4.35524954951622 0.143778372575268
4.35841997001659 0.170244456816511
4.36159039051695 0.200843227615254
4.36476081101732 0.236072934844829
4.36793123151768 0.276464900360839
4.37110165201805 0.32258086297689
4.37427207251841 0.375009260596227
4.37744249301878 0.434360356404465
4.38061291351914 0.501260132350176
4.38378333401951 0.576342894603564
4.38695375451987 0.660242562370592
4.39012417502024 0.753582643240854
4.3932945955206 0.856964934833564
4.39646501602097 0.970957033310708
4.39963543652134 1.0960787735333
4.4028058570217 1.23278777217752
4.40597627752207 1.3814642926944
4.40914669802243 1.5423956980579
4.4123171185228 1.71576080209543
4.41548753902316 1.9016144709815
4.41865795952353 2.09987286128549
4.42182838002389 2.31029970787936
4.42499880052426 2.5324940921969
4.42816922102462 2.7658801271254
4.43133964152499 3.00969898779668
4.43451006202535 3.26300369666362
4.43768048252572 3.52465703586171
4.44085090302608 3.79333290982064
4.44402132352645 4.06752141680225
4.44719174402681 4.34553781048589
4.45036216452718 4.62553544345631
4.45353258502754 4.90552268561157
4.45670300552791 5.18338370475588
4.45987342602827 5.45690288708667
4.46304384652864 5.72379256539229
4.466214267029 5.98172361625076
4.46938468752937 6.22835838815748
4.47255510802973 6.46138533405777
4.4757255285301 6.67855464775127
4.47889594903046 6.8777141472326
4.48206636953083 7.05684461189273
4.48523679003119 7.21409376662747
4.48840721053156 7.34780811554017
4.49157763103192 7.45656186150507
4.49474805153229 7.53918220492418
4.49791847203266 7.59477039423177
4.50108889253302 7.62271799989771
4.50425931303339 7.62271799989771
4.50742973353375 7.59477039423177
4.51060015403412 7.53918220492418
4.51377057453448 7.45656186150507
4.51694099503485 7.34780811554017
4.52011141553521 7.21409376662747
4.52328183603558 7.05684461189273
4.52645225653594 6.8777141472326
4.52962267703631 6.67855464775127
4.53279309753667 6.46138533405777
4.53596351803704 6.22835838815748
4.5391339385374 5.98172361625076
4.54230435903777 5.72379256539229
4.54547477953813 5.45690288708667
4.5486452000385 5.18338370475588
4.55181562053886 4.90552268561157
4.55498604103923 4.62553544345631
4.55815646153959 4.34553781048589
4.56132688203996 4.06752141680225
4.56449730254032 3.79333290982064
4.56766772304069 3.52465703586171
4.57083814354105 3.26300369666362
4.57400856404142 3.00969898779668
4.57717898454178 2.7658801271254
4.58034940504215 2.5324940921969
4.58351982554252 2.31029970787936
4.58669024604288 2.09987286128549
4.58986066654325 1.9016144709815
4.59303108704361 1.71576080209543
4.59620150754398 1.5423956980579
4.59937192804434 1.3814642926944
4.60254234854471 1.23278777217752
4.60571276904507 1.0960787735333
4.60888318954544 0.970957033310708
4.6120536100458 0.856964934833564
4.61522403054617 0.753582643240854
4.61839445104653 0.660242562370592
4.6215648715469 0.576342894603564
4.62473529204726 0.501260132350176
4.62790571254763 0.434360356404465
4.63107613304799 0.375009260596227
4.63424655354836 0.32258086297689
4.63741697404872 0.276464900360839
4.64058739454909 0.236072934844829
4.64375781504945 0.200843227615254
4.64692823554982 0.170244456816511
4.65009865605018 0.143778372575268
4.65326907655055 0.120981493700508
4.65643949705091 0.10142595748068
4.65960991755128 0.0847196368493868
};
\end{axis}

\end{tikzpicture}

\medskip

\begin{tikzpicture}

\begin{axis}[
width=3.0in,
height=1.5in,
axis lines=left,
enlarge x limits=0.1,
ylabel={Density},
ylabel near ticks,
ytick=\empty,
legend entries={{true $\beta_2$}, {$q(\beta_2)$}},
legend style={draw=none, at={(1.0,1.0)}, anchor=north east,font=\small},
legend cell align=left,
]
\addplot [thick, dashed, gray]
table {%
2.31993941811405 0
2.31993941811405 10
};
\addplot [very thick, MidnightBlue]
table {%
2.17749552056193 0.0994200051212747
2.18019715865905 0.119025170399235
2.18289879675616 0.141973940996314
2.18560043485327 0.168726485019929
2.18830207295039 0.199784906994549
2.1910037110475 0.235693110365672
2.19370534914461 0.277035800247729
2.19640698724172 0.324436492316309
2.19910862533884 0.378554397089809
2.20181026343595 0.440080056944581
2.20451190153306 0.509729626615162
2.20721353963017 0.588237707084861
2.20991517772729 0.676348667959524
2.2126168158244 0.774806424735553
2.21531845392151 0.884342674691867
2.21802009201863 1.00566363806981
2.22072173011574 1.13943539909057
2.22342336821285 1.28626799323666
2.22612500630996 1.44669844184086
2.22882664440708 1.62117299084632
2.23152828250419 1.8100288658291
2.2342299206013 2.01347590800542
2.23693155869842 2.23157850380988
2.23963319679553 2.46423826148096
2.24233483489264 2.71117791967622
2.24503647298975 2.97192699330649
2.24773811108687 3.24580966857194
2.25043974918398 3.53193545095336
2.25314138728109 3.82919304540647
2.25584302537821 4.13624790648096
2.25854466347532 4.45154383736945
2.26124630157243 4.77330894144763
2.26394793966954 5.09956613885468
2.26664957776666 5.42814835590733
2.26935121586377 5.75671837915301
2.27205285396088 6.0827932417665
2.274754492058 6.40377288142713
2.27745613015511 6.71697267985326
2.28015776825222 7.01965936916085
2.28285940634933 7.3090896736089
2.28556104444645 7.58255095149406
2.28826268254356 7.83740301510686
2.29096432064067 8.07112024047207
2.29366595873779 8.2813330361849
2.2963675968349 8.46586772436868
2.29906923493201 8.62278389809264
2.30177087302912 8.75040835899549
2.30447251112624 8.84736480582791
2.30717414922335 8.91259853759506
2.30987578732046 8.94539555138979
2.31257742541758 8.94539555138979
2.31527906351469 8.91259853759505
2.3179807016118 8.84736480582791
2.32068233970891 8.75040835899549
2.32338397780603 8.62278389809264
2.32608561590314 8.46586772436868
2.32878725400025 8.2813330361849
2.33148889209736 8.07112024047207
2.33419053019448 7.83740301510686
2.33689216829159 7.58255095149406
2.3395938063887 7.3090896736089
2.34229544448582 7.01965936916085
2.34499708258293 6.71697267985321
2.34769872068004 6.40377288142713
2.35040035877715 6.0827932417665
2.35310199687427 5.75671837915301
2.35580363497138 5.42814835590733
2.35850527306849 5.09956613885468
2.36120691116561 4.77330894144763
2.36390854926272 4.45154383736945
2.36661018735983 4.13624790648096
2.36931182545694 3.82919304540647
2.37201346355406 3.53193545095336
2.37471510165117 3.24580966857189
2.37741673974828 2.97192699330649
2.3801183778454 2.71117791967622
2.38282001594251 2.46423826148096
2.38552165403962 2.23157850380988
2.38822329213673 2.01347590800542
2.39092493023385 1.8100288658291
2.39362656833096 1.62117299084632
2.39632820642807 1.44669844184086
2.39902984452519 1.28626799323666
2.4017314826223 1.13943539909057
2.40443312071941 1.00566363806979
2.40713475881652 0.884342674691867
2.40983639691364 0.774806424735553
2.41253803501075 0.676348667959524
2.41523967310786 0.588237707084861
2.41794131120498 0.509729626615162
2.42064294930209 0.440080056944581
2.4233445873992 0.378554397089809
2.42604622549631 0.324436492316309
2.42874786359343 0.277035800247729
2.43144950169054 0.235693110365672
2.43415113978765 0.199784906994543
2.43685277788477 0.168726485019929
2.43955441598188 0.141973940996314
2.44225605407899 0.119025170399235
2.4449576921761 0.0994200051212747
};
\end{axis}

\end{tikzpicture}
\begin{tikzpicture}

\begin{axis}[
width=3.0in,
height=1.5in,
axis lines=left,
enlarge x limits=0.1,
ylabel={Density},
ylabel near ticks,
ytick=\empty,
legend entries={{true $\beta_3$}, {$q(\beta_3)$}},
legend style={draw=none, at={(1.0,1.0)}, anchor=north east,font=\small},
legend cell align=left,
]
\addplot [thick, dashed, gray]
table {%
0.986584841970366 0
0.986584841970366 10
};
\addplot [very thick, MidnightBlue]
table {%
0.845995739102364 0.0957297390446682
0.848801522092386 0.114607200916625
0.851607305082408 0.136704160356275
0.85441308807243 0.162463704977476
0.857218871062452 0.192369302217613
0.860024654052474 0.226944666945143
0.862830437042496 0.26675280122341
0.865636220032519 0.312394077108754
0.868442003022541 0.364503236581151
0.871247786012563 0.423745190504442
0.874053569002585 0.49080951142261
0.876859351992607 0.566403533441557
0.879665134982629 0.651243996698592
0.882470917972651 0.746047204077195
0.885276700962674 0.851517693758399
0.888082483952696 0.968335472540955
0.890888266942718 1.09714190096996
0.89369404993274 1.23852437126565
0.896499832922762 1.39299997163367
0.899305615912784 1.56099838428567
0.902111398902806 1.74284431767804
0.904917181892829 1.93873982415276
0.907722964882851 2.14874690025235
0.910528747872873 2.37277080631509
0.913334530862895 2.61054457236931
0.916140313852917 2.86161517676441
0.918946096842939 3.12533189051869
0.921751879832961 3.40083727243771
0.924557662822983 3.68706127646313
0.927363445813006 3.98271889272663
0.930169228803028 4.28631168724502
0.93297501179305 4.59613353255179
0.935780794783072 4.91028073392366
0.938586577773094 5.22666665499454
0.941392360763116 5.54304083486749
0.944198143753138 5.85701246933826
0.947003926743161 6.16607800505092
0.949809709733183 6.46765247123206
0.952615492723205 6.75910405327903
0.955421275713227 7.03779130020314
0.958227058703249 7.30110225798107
0.961032841693271 7.54649473724003
0.963838624683293 7.77153685997138
0.966644407673316 7.97394698912894
0.969450190663338 8.15163212928732
0.97225597365336 8.30272389742954
0.975061756643382 8.4256112008756
0.977867539633404 8.5189688238474
0.980673322623426 8.58178121368143
0.983479105613448 8.61336086978972
0.986284888603471 8.61336086978972
0.989090671593493 8.58178121368143
0.991896454583515 8.5189688238474
0.994702237573537 8.4256112008756
0.997508020563559 8.30272389742954
1.00031380355358 8.15163212928732
1.0031195865436 7.97394698912894
1.00592536953363 7.77153685997139
1.00873115252365 7.54649473724004
1.01153693551367 7.30110225798107
1.01434271850369 7.03779130020314
1.01714850149371 6.75910405327903
1.01995428448374 6.46765247123206
1.02276006747376 6.16607800505091
1.02556585046378 5.85701246933825
1.0283716334538 5.54304083486749
1.03117741644382 5.22666665499454
1.03398319943385 4.91028073392366
1.03678898242387 4.5961335325518
1.03959476541389 4.28631168724503
1.04240054840391 3.98271889272663
1.04520633139394 3.68706127646313
1.04801211438396 3.40083727243771
1.05081789737398 3.12533189051869
1.053623680364 2.8616151767644
1.05642946335402 2.6105445723693
1.05923524634405 2.37277080631509
1.06204102933407 2.14874690025235
1.06484681232409 1.93873982415276
1.06765259531411 1.74284431767805
1.07045837830413 1.56099838428568
1.07326416129416 1.39299997163367
1.07606994428418 1.23852437126565
1.0788757272742 1.09714190096996
1.08168151026422 0.96833547254096
1.08448729325425 0.851517693758395
1.08729307624427 0.746047204077192
1.09009885923429 0.651243996698592
1.09290464222431 0.566403533441557
1.09571042521433 0.49080951142261
1.09851620820436 0.423745190504444
1.10132199119438 0.364503236581153
1.1041277741844 0.312394077108753
1.10693355717442 0.26675280122341
1.10973934016444 0.226944666945143
1.11254512315447 0.192369302217614
1.11535090614449 0.162463704977475
1.11815668913451 0.136704160356274
1.12096247212453 0.114607200916625
1.12376825511456 0.0957297390446682
};
\end{axis}

\end{tikzpicture}

\medskip

\begin{tikzpicture}

\begin{axis}[
width=3.0in,
height=1.5in,
axis lines=left,
enlarge x limits=0.1,
ylabel={Density},
ylabel near ticks,
ytick=\empty,
legend entries={{true $\beta_4$}, {$q(\beta_4)$}},
legend style={draw=none, at={(1.0,1.0)}, anchor=north east,font=\small},
legend cell align=left,
]
\addplot [thick, dashed, gray]
table {%
-3.43981359557564 0
-3.43981359557564 10
};
\addplot [very thick, MidnightBlue]
table {%
-3.56477700546384 0.0990877818607054
-3.56206630926692 0.118627434247927
-3.55935561307 0.141499518916615
-3.55664491687309 0.168162666270076
-3.55393422067617 0.199117302993386
-3.55122352447925 0.234905515016792
-3.54851282828233 0.276110053596034
-3.54580213208542 0.323352350858125
-3.5430914358885 0.3772894144945
-3.54038073969158 0.43860947935558
-3.53767004349467 0.508026307063375
-3.53495934729775 0.586272043848377
-3.53224865110083 0.674088571920725
-3.52953795490391 0.772217320898544
-3.526827258707 0.881387543010997
-3.52411656251008 1.00230309858413
-3.52140586631316 1.13562784604295
-3.51869517011624 1.28196978236695
-3.51598447391933 1.44186413437105
-3.51327377772241 1.61575565681672
-3.51056308152549 1.80398044840246
-3.50785238532857 2.00674764913621
-3.50514168913166 2.2241214302979
-3.50243099293474 2.45600372891326
-3.49972029673782 2.70211821014031
-3.49700960054091 2.96199596106921
-3.49429890434399 3.23496342620889
-3.49158820814707 3.52013308672892
-3.48887751195015 3.81639736110385
-3.48616681575324 4.12242616341894
-3.48345611955632 4.43666849707539
-3.4807454233594 4.75735838644311
-3.47803472716248 5.08252535829877
-3.47532403096557 5.41000958048226
-3.47261333476865 5.73748164960578
-3.46990263857173 6.06246689596034
-3.46719194237481 6.38237394562925
-3.4644812461779 6.6945271512885
-3.46177054998098 6.99620237857971
-3.45905985378406 7.28466551872985
-3.45634915758715 7.55721299463653
-3.45363846139023 7.81121344107799
-3.45092776519331 8.04414967373889
-3.44821706899639 8.2536600194738
-3.44550637279948 8.43757806399793
-3.44279567660256 8.59396988447118
-3.44008498040564 8.72116787371497
-3.43737428420872 8.81780032954718
-3.43466358801181 8.88281607537638
-3.43195289181489 8.9155034942174
-3.42924219561797 8.9155034942174
-3.42653149942105 8.88281607537638
-3.42382080322414 8.81780032954718
-3.42111010702722 8.72116787371497
-3.4183994108303 8.59396988447118
-3.41568871463339 8.43757806399793
-3.41297801843647 8.2536600194738
-3.41026732223955 8.04414967373889
-3.40755662604263 7.81121344107799
-3.40484592984572 7.55721299463653
-3.4021352336488 7.2846655187298
-3.39942453745188 6.99620237857971
-3.39671384125496 6.6945271512885
-3.39400314505805 6.38237394562925
-3.39129244886113 6.06246689596034
-3.38858175266421 5.73748164960578
-3.38587105646729 5.41000958048226
-3.38316036027038 5.08252535829877
-3.38044966407346 4.75735838644311
-3.37773896787654 4.43666849707539
-3.37502827167963 4.12242616341894
-3.37231757548271 3.81639736110385
-3.36960687928579 3.52013308672892
-3.36689618308887 3.23496342620889
-3.36418548689196 2.96199596106921
-3.36147479069504 2.70211821014031
-3.35876409449812 2.45600372891326
-3.3560533983012 2.2241214302979
-3.35334270210429 2.00674764913621
-3.35063200590737 1.80398044840246
-3.34792130971045 1.61575565681672
-3.34521061351353 1.44186413437105
-3.34249991731662 1.28196978236695
-3.3397892211197 1.13562784604295
-3.33707852492278 1.00230309858413
-3.33436782872587 0.881387543010997
-3.33165713252895 0.772217320898544
-3.32894643633203 0.674088571920725
-3.32623574013511 0.586272043848377
-3.3235250439382 0.508026307063375
-3.32081434774128 0.43860947935558
-3.31810365154436 0.3772894144945
-3.31539295534744 0.323352350858125
-3.31268225915053 0.276110053596027
-3.30997156295361 0.234905515016792
-3.30726086675669 0.199117302993386
-3.30455017055977 0.168162666270076
-3.30183947436286 0.141499518916615
-3.29912877816594 0.118627434247927
-3.29641808196902 0.0990877818607054
};
\end{axis}

\end{tikzpicture}
\caption{Visualization of the inferred marginal posteriors for Bayesian linear
regression. The gray bars indicate the simulated ``true'' value for each
component of the coefficient vector.}
\label{fig:supervised-betas}
\end{figure}

Figure\nobreakspace \ref{fig:supervised-betas} shows the resulting posteriors from  variational
inference. We plot the marginal posteriors for each component of the vector of
coefficients $\beta$. The vertical lines indicate the ``true'' values of the
coefficients that we simulated above.

\subsubsection{Criticism}

A standard evaluation in regression is to calculate point-based evaluations on
held-out ``testing'' data. We do this first by forming the posterior predictive
distribution.
\begin{lstlisting}
y_post = Normal(mu=ed.dot(X, qw.mean()) + qb.mean(), sigma=tf.ones(N))
\end{lstlisting}

With this we can evaluate various point-based quantities using the posterior
predictive.
\begin{lstlisting}
print(ed.evaluate('mean_squared_error', data={X: X_test, y_post: y_test}))
> 0.012107

print(ed.evaluate('mean_absolute_error', data={X: X_test, y_post: y_test}))
> 0.0867875
\end{lstlisting}

The trained model makes predictions with low mean squared error
(relative to the magnitude of the output).

Edward supports another class of criticism techniques called
\glspl{PPC}.
The simplest \gls{PPC} works by applying a test statistic on new data
generated from the posterior predictive, such as
$T(\mathbf{x}_\text{new}) = \max(\mathbf{x}_\text{new})$.
Applying $T(\mathbf{x}_\text{new})$ to
new data over many data replications induces a distribution of the test
statistic, $\textsc{ppd}(T)$. We compare
this distribution to the test statistic applied to the original dataset
$T(\mathbf{x})$.

\begin{figure}[htb]
\centering
\begin{tikzpicture}

\begin{axis}[
width=3.0in,
height=1.5in,
axis lines=left,
enlarge x limits=0.1,
ylabel={Density},
ylabel near ticks,
ytick=\empty,
legend entries={{$T_\text{min}(\mathbf{x})$}, {$\textsc{ppd}(T_\text{min})$}},
legend style={draw=none, at={(1.0,1.0)}, anchor=north east,font=\small},
legend cell align=left,
]
\addplot [thick, dashed, gray]
table {%
-16.5441242364738 0
-16.5441242364738 0.7
};
\addplot [very thick, MidnightBlue]
table {%
-13.2352993891791 5.05219293070858e-13
-13.3021443355891 2.05825068882034e-12
-13.3689892819991 8.038769482291e-12
-13.4358342284091 3.01002711420105e-11
-13.5026791748191 1.08058842088904e-10
-13.5695241212291 3.71948112238135e-10
-13.6363690676391 1.22761967091541e-09
-13.703214014049 3.88541167689218e-09
-13.770058960459 1.17934327544065e-08
-13.836903906869 3.43335601433984e-08
-13.903748853279 9.58797378067529e-08
-13.970593799689 2.56878942376166e-07
-14.037438746099 6.60390007322067e-07
-14.104283692509 1.62942503466722e-06
-14.171128638919 3.85957766174606e-06
-14.237973585329 8.77906132307628e-06
-14.304818531739 1.91830211203899e-05
-14.371663478149 4.02843232474005e-05
-14.438508424559 8.13454163652436e-05
-14.505353370969 0.000158045340654349
-14.572198317379 0.000295671862401375
-14.639043263789 0.000533101503267019
-14.705888210199 0.000927356629459576
-14.772733156609 0.00155836143375224
-14.839578103019 0.0025334377221303
-14.906423049429 0.00399115809416706
-14.973267995839 0.00610441074410488
-15.040112942249 0.00908280448773802
-15.1069578886589 0.0131746032563921
-15.1738028350689 0.0186679352142648
-15.2406477814789 0.0258899291011952
-15.3074927278889 0.0352009076844701
-15.3743376742989 0.0469795037322437
-15.4411826207089 0.061594575989866
-15.5080275671189 0.0793620014551975
-15.5748725135289 0.100488999535781
-15.6417174599389 0.125014606287185
-15.7085624063489 0.152760090791724
-15.7754073527589 0.183304841215112
-15.8422522991689 0.215999631691617
-15.9090972455789 0.250020225923821
-15.9759421919889 0.284452402130418
-16.0427871383989 0.318388874103353
-16.1096320848089 0.351013519104097
-16.1764770312189 0.38165129488892
-16.2433219776289 0.409772728951764
-16.3101669240389 0.434956507091933
-16.3770118704489 0.456827556412376
-16.4438568168589 0.474996523312243
-16.5107017632689 0.489026959953148
-16.5775467096788 0.498448653141407
-16.6443916560888 0.502821529311249
-16.7112366024988 0.501838316951246
-16.7780815489088 0.495440274532375
-16.8449264953188 0.483913062864297
-16.9117714417288 0.467931894223707
-16.9786163881388 0.448536362773748
-17.0454613345488 0.427032755011925
-17.1123062809588 0.404839873801779
-17.1791512273688 0.383307916393971
-17.2459961737788 0.363545023299929
-17.3128411201888 0.346282271150708
-17.3796860665988 0.331797694426288
-17.4465310130088 0.319907513582849
-17.5133759594188 0.31002177463391
-17.5802209058288 0.301253806284686
-17.6470658522388 0.292567919168833
-17.7139107986488 0.282946187091122
-17.7807557450588 0.271552000520565
-17.8476006914688 0.257865922265467
-17.9144456378788 0.241770195825571
-17.9812905842888 0.223564258665736
-18.0481355306987 0.20390550471423
-18.1149804771087 0.183685389894948
-18.1818254235187 0.163866398952204
-18.2486703699287 0.145315174953473
-18.3155153163387 0.128667506515823
-18.3823602627487 0.114251399827959
-18.4492052091587 0.102078285357883
-18.5160501555687 0.0918949883808782
-18.5828951019787 0.0832759071210482
-18.6497400483887 0.0757292467898767
-18.7165849947987 0.0687933906826522
-18.7834299412087 0.0621071419943456
-18.8502748876187 0.0554470682033875
-18.9171198340287 0.0487334112562898
-18.9839647804387 0.0420113379764541
-19.0508097268487 0.0354165758884345
-19.1176546732587 0.0291345116183055
-19.1844996196687 0.023360572096513
-19.2513445660787 0.0182677812430854
-19.3181895124887 0.0139850418440157
-19.3850344588987 0.0105870889772692
-19.4518794053086 0.00809451956674418
-19.5187243517186 0.0064803640817207
-19.5855692981286 0.0056788992121118
-19.6524142445386 0.00559309392058319
-19.7192591909486 0.00609902076360903
-19.7861041373586 0.00704802346396004
-19.8529490837686 0.00826940904339173
};
\end{axis}

\end{tikzpicture}
\begin{tikzpicture}

\begin{axis}[
width=3.0in,
height=1.5in,
axis lines=left,
enlarge x limits=0.1,
ylabel={Density},
ylabel near ticks,
ytick=\empty,
legend entries={{$T_\text{max}(\mathbf{x})$}, {$\textsc{ppd}(T_\text{max})$}},
legend style={draw=none, at={(1.0,1.0)}, anchor=north east,font=\small},
legend cell align=left,
]
\addplot [thick, dashed, gray]
table {%
16.6090179263646 0
16.6090179263646 0.7
};
\addplot [very thick, MidnightBlue]
table {%
13.2872143410917 5.88907908239748e-12
13.3543214842285 2.30682665294251e-11
13.4214286273653 8.65546789880128e-11
13.4885357705022 3.1109455968171e-10
13.555642913639 1.0711268332124e-09
13.6227500567758 3.53312245190288e-09
13.6898571999126 1.11653193225071e-08
13.7569643430495 3.3807003201652e-08
13.8240714861863 9.80837688954409e-08
13.8911786293231 2.72695790757017e-07
13.9582857724599 7.26592725735176e-07
14.0253929155968 1.85558514831365e-06
14.0925000587336 4.54255547295968e-06
14.1596072018704 1.06612705924626e-05
14.2267143450072 2.39925316082358e-05
14.2938214881441 5.17823461453552e-05
14.3609286312809 0.0001072061804853
14.4280357744177 0.000212961676966936
14.4951429175546 0.000406031146985154
14.5622500606914 0.000743282168077874
14.6293572038282 0.00130701292524395
14.696464346965 0.0022089106794077
14.7635714901019 0.00359042894948312
14.8306786332387 0.00561765895060876
14.8977857763755 0.00846970343296913
14.9648929195123 0.0123214421175529
15.0320000626492 0.0173240355812697
15.099107205786 0.0235886564903373
15.1662143489228 0.031179576611542
15.2333214920596 0.0401209248561671
15.3004286351965 0.0504171026374971
15.3675357783333 0.0620812265945392
15.4346429214701 0.0751613769177472
15.501750064607 0.0897533274782765
15.5688572077438 0.105992124712952
15.6359643508806 0.124022507817905
15.7030714940174 0.143956727359728
15.7701786371543 0.165833949246076
15.8372857802911 0.189595328748755
15.9043929234279 0.215082843128724
15.9715000665647 0.242060737082459
16.0386072097016 0.270250050733925
16.1057143528384 0.299362541219018
16.1728214959752 0.32912144615983
16.239928639112 0.359261610668238
16.3070357822489 0.389507960577613
16.3741429253857 0.419537018754109
16.4412500685225 0.448930257716725
16.5083572116594 0.477130658103593
16.5754643547962 0.503415058089259
16.642571497933 0.526894349720654
16.7096786410698 0.546550574825499
16.7767857842067 0.561314151873328
16.8438929273435 0.570176434087392
16.9110000704803 0.572324061977174
16.9781072136171 0.567274001154524
17.045214356754 0.554983501388671
17.1123214998908 0.535908974244704
17.1794286430276 0.51099320138505
17.2465357861644 0.481571955357454
17.3136429293013 0.449208111466936
17.3807500724381 0.415480326874349
17.4478572155749 0.381768753084954
17.5149643587118 0.349085826732023
17.5820715018486 0.317991959931194
17.6491786449854 0.28861464804705
17.7162857881222 0.260761178662522
17.7833929312591 0.234089181792312
17.8505000743959 0.208284730263077
17.9176072175327 0.183199427299166
17.9847143606695 0.15891462704636
18.0518215038064 0.135725953354043
18.1189286469432 0.114065475069874
18.18603579008 0.0943943177157276
18.2531429332168 0.0771011977617766
18.3202500763537 0.062433366545238
18.3873572194905 0.0504707003030209
18.4544643626273 0.0411377501446338
18.5215715057642 0.0342380739875309
18.588678648901 0.0294928336595045
18.6557857920378 0.0265706969754408
18.7228929351746 0.025105076939142
18.7900000783115 0.0247032169000523
18.8571072214483 0.0249560346592097
18.9242143645851 0.0254566143186622
18.9913215077219 0.0258298999634965
19.0584286508588 0.0257693361585591
19.1255357939956 0.0250711059829185
19.1926429371324 0.0236554243582867
19.2597500802692 0.0215674372259132
19.3268572234061 0.0189562594568207
19.3939643665429 0.0160371232920431
19.4610715096797 0.0130461174367693
19.5281786528165 0.0101981696569966
19.5952857959534 0.00765675999972511
19.6623929390902 0.00551954424981433
19.729500082227 0.0038193648300069
19.7966072253639 0.00253657419000109
19.8637143685007 0.00161701386050297
19.9308215116375 0.000990349592964849
};
\end{axis}

\end{tikzpicture}

\medskip

\begin{tikzpicture}

\begin{axis}[
width=3.0in,
height=1.5in,
axis lines=left,
enlarge x limits=0.1,
ylabel={Density},
ylabel near ticks,
ytick=\empty,
legend entries={{$T_\text{mean}(\mathbf{x})$}, {$\textsc{ppd}(T_\text{mean})$}},
legend style={draw=none, at={(1.0,1.0)}, anchor=north east,font=\small},
legend cell align=left,
]
\addplot [thick, dashed, gray]
table {%
-0.559917133751622 0
-0.559917133751622 8
};
\addplot [very thick, MidnightBlue]
table {%
-0.335950280250973 1.31826193247141e-07
-0.340474863149976 4.40417152547664e-07
-0.344999446048979 1.40051041254e-06
-0.349524028947982 4.23967782308379e-06
-0.354048611846985 1.22201874775514e-05
-0.358573194745988 3.3543340518437e-05
-0.363097777644991 8.77032514473706e-05
-0.367622360543994 0.000218483251802879
-0.372146943442997 0.000518731692386615
-0.376671526342 0.0011741899396429
-0.381196109241003 0.00253499069727522
-0.385720692140006 0.00522221230327025
-0.390245275039009 0.0102707121664167
-0.394769857938012 0.019296407749699
-0.399294440837015 0.0346563621203797
-0.403819023736018 0.0595475810788749
-0.408343606635021 0.097974212874276
-0.412868189534024 0.154514642298559
-0.417392772433027 0.233849438781403
-0.42191735533203 0.34006905738358
-0.426441938231033 0.475853928256519
-0.430966521130036 0.641684585993456
-0.435491104029039 0.835268781763808
-0.440015686928042 1.05135004425149
-0.444540269827045 1.28199255463699
-0.449064852726048 1.51734460173359
-0.453589435625051 1.74679787721112
-0.458114018524054 1.96040229159795
-0.462638601423057 2.15036528409097
-0.46716318432206 2.31244527582332
-0.471687767221063 2.4470278702521
-0.476212350120066 2.55965637579346
-0.480736933019069 2.66080216636917
-0.485261515918072 2.76473802573947
-0.489786098817075 2.88753333272035
-0.494310681716078 3.04440269712009
-0.498835264615081 3.24685395853415
-0.503359847514084 3.50022588193761
-0.507884430413087 3.80221894074062
-0.51240901331209 4.14287445862823
-0.516933596211093 4.5061614945854
-0.521458179110096 4.87294490313373
-0.525982762009099 5.22472548384073
-0.530507344908102 5.54727438361957
-0.535031927807105 5.83322641691359
-0.539556510706108 6.08290049624787
-0.544081093605111 6.30305460135884
-0.548605676504114 6.50385157623385
-0.553130259403117 6.69484690038952
-0.55765484230212 6.88113690254679
-0.562179425201123 7.06080036809693
-0.566704008100126 7.22440154665388
-0.571228590999129 7.35669463994601
-0.575753173898132 7.43998095855344
-0.580277756797135 7.4580631623682
-0.584802339696138 7.39960494788908
-0.589326922595141 7.25998696678778
-0.593851505494144 7.04132844890974
-0.598376088393147 6.75097937727867
-0.60290067129215 6.39923484102435
-0.607425254191153 5.99714012155065
-0.611949837090156 5.55505271572972
-0.616474419989159 5.08223599892622
-0.620999002888162 4.58734843736424
-0.625523585787165 4.07939787597822
-0.630048168686168 3.56862012110047
-0.634572751585171 3.06681792580805
-0.639097334484174 2.58691662728942
-0.643621917383177 2.14178028998728
-0.64814650028218 1.74259488341976
-0.652671083181183 1.39727785944853
-0.657195666080186 1.10936936894104
-0.661720248979189 0.87770989538598
-0.666244831878192 0.696974855475188
-0.670769414777195 0.558903275250046
-0.675293997676198 0.453895730432368
-0.679818580575201 0.372601409041363
-0.684343163474204 0.307163343307327
-0.688867746373207 0.251916462815002
-0.69339232927221 0.203493603930414
-0.697916912171213 0.160444778103095
-0.702441495070216 0.12257535509535
-0.706966077969219 0.0902362521484529
-0.711490660868222 0.0637553138982275
-0.716015243767225 0.0431095405530541
-0.720539826666228 0.0278410195212284
-0.725064409565231 0.0171492513326823
-0.729588992464234 0.0100653399100707
-0.734113575363237 0.00562517074090984
-0.73863815826224 0.00299195464813669
-0.743162741161243 0.00151403282434576
-0.747687324060246 0.000728732711755793
-0.752211906959249 0.000333561443670076
-0.756736489858252 0.000145178214655598
-0.761261072757255 6.00764012086311e-05
-0.765785655656258 2.363486586849e-05
-0.770310238555261 8.83948805819745e-06
-0.774834821454264 3.14274855194534e-06
-0.779359404353267 1.0621571788719e-06
-0.78388398725227 3.41236740996593e-07
};
\end{axis}

\end{tikzpicture}
\glsreset{PPC}
\caption{Examples of \glspl{PPC} for Bayesian linear regression. }
\label{fig:supervised-ppc}
\end{figure}
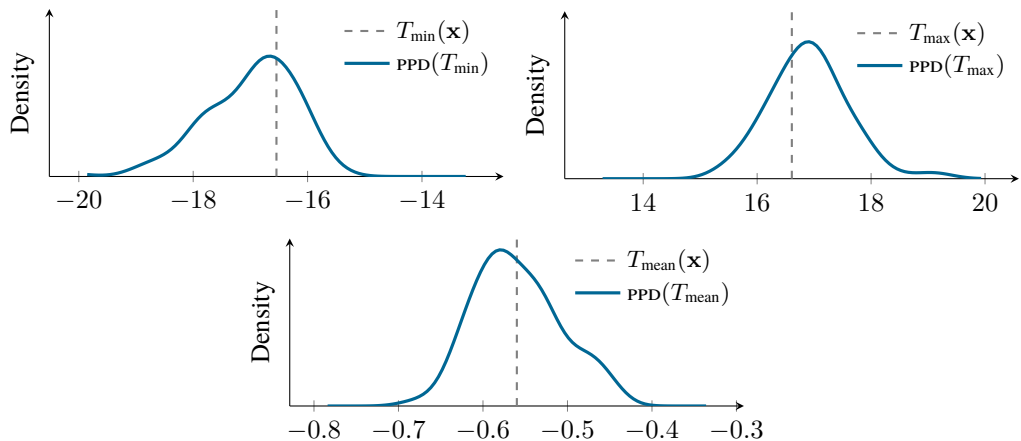

Calculating \glspl{PPC} in Edward is straightforward.
\begin{lstlisting}
def T(xs, zs):
  return tf.reduce_max(xs[y_post])

ppc_max = ed.ppc(T, data={X: X_train, y_post: y_train})
\end{lstlisting}
This calculates the test statistic on both the original dataset as well as on
data replications generated from teh posterior predictive distribution.
Figure \ref{fig:supervised-ppc} shows three visualizations of different \glspl
{PPC}; the
plotted posterior predictive distributions are kernel density estimates from
$N=500$ data replications.

\subsection{Logistic and Neural Network Classification}

In supervised learning, the task is to infer hidden structure from
labeled data, comprised of training examples $\{(x_n, y_n)\}$.
Classification means the output $y$ takes discrete values.

\subsubsection{Data}

We study a two-dimensional simulated dataset with a nonlinear decision
boundary. We simulate $100$ datapoints using the following snippet.
\begin{lstlisting}
from scipy.stats import logistic

N = 100  # number of data points
D = 2  # number of features

px1 = np.linspace(-3, 3, 50)
px2 = np.linspace(-3, 3, 50)
px1_m, px2_m = np.mgrid[-3:3:50j, -3:3:50j]

xeval = np.vstack((px1_m.flatten(), px2_m.flatten())).T
x_viz = tf.constant(np.array(xeval, dtype='float32'))

def build_toy_dataset(N):
  x = xeval[np.random.randint(xeval.shape[0],size=N), :]
  y = bernoulli.rvs(p=logistic.cdf( 5 * x[:, 0]**2 + 5 * x[:, 1]**3 ))
  return x, y

x_train, y_train = build_toy_dataset(N)
\end{lstlisting}

\begin{figure}[!htbp]
\centering
\includegraphics[width=2.5in]{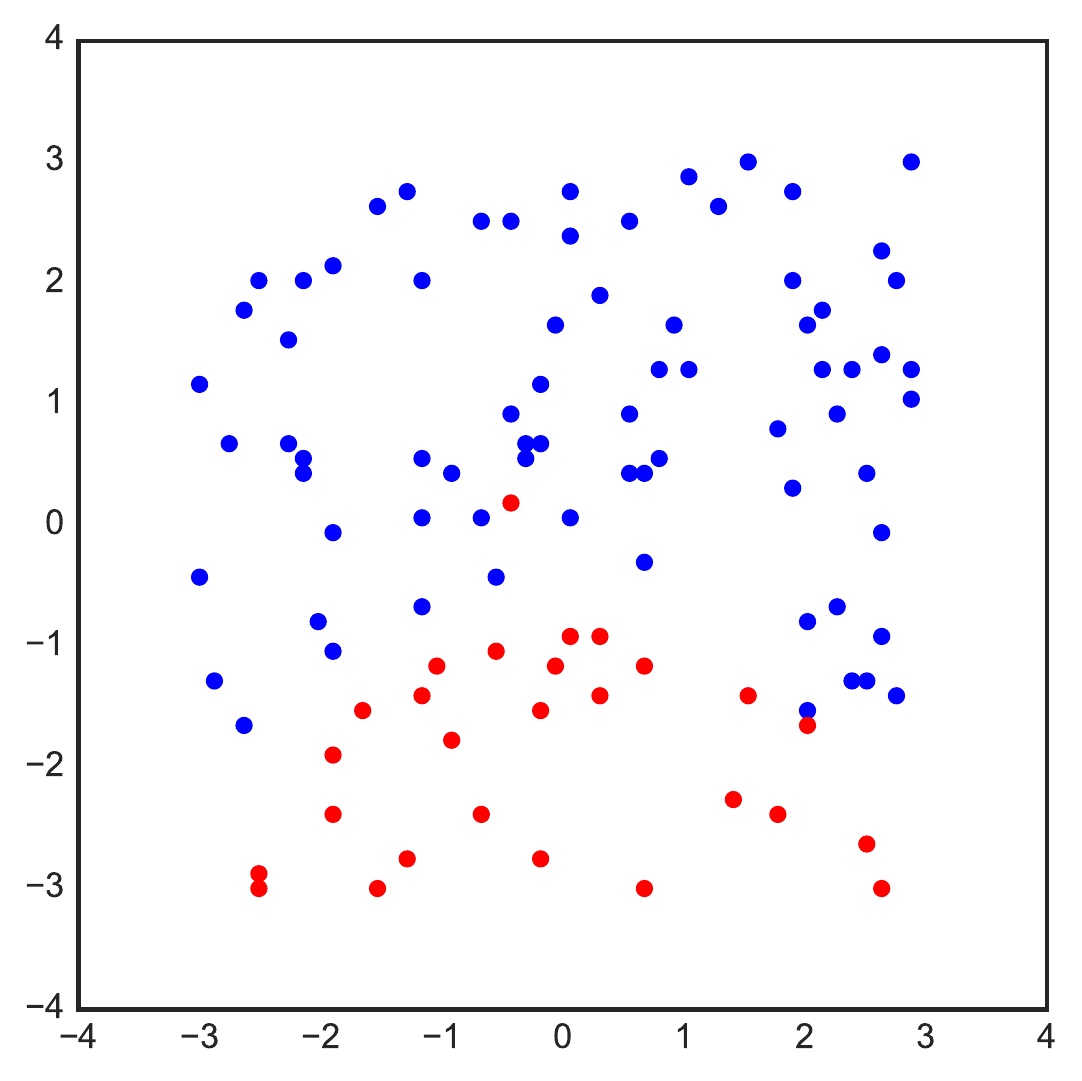}
\caption{Simulated data for classification. Positive and negative measurements
colored by label.}
\label{fig:lr_data}
\end{figure}

Figure\nobreakspace \ref{fig:lr_data} shows the data, colored by label. The red point near the
origin makes this a challenging dataset for classification models that assume a
linear decision boundary.

\subsubsection{Model: Bayesian Logistic Regression}

We begin with a popular classification model: logistic regression. This model
relates outputs $y\in\{0,1\}$, also known as the response, given
a vector of inputs $\mathbf{x}\in\mathbb{R}^D$, also known as the features or
covariates. The model assumes a latent linear relationship between these two random
variables \citep{gelman2013bayesian}.

The likelihood of each datapoint is a Bernoulli with probability
\begin{align*}
\Pr(y_n=1)
  &=
  \text{logistic}
  \left(
  \mathbf{x}^\top \mathbf{w} + b
  \right).
\end{align*}
We posit priors on the latent variables $\mathbf{w}$ and $b$ as
\begin{align*}
  p(\mathbf{w})
  &=
  \text{Normal}(\mathbf{w} \mid \mathbf{0}, \sigma_w^2\mathbf{I}),
  \\
  p(b)
  &=
  \text{Normal}(b \mid 0, \sigma_b^2).
\end{align*}

This model is
easy to specify in Edward's native language.

\begin{lstlisting}
W = Normal(mu=tf.zeros(D), sigma=tf.ones(D))
b = Normal(mu=tf.zeros(1), sigma=tf.ones(1))

x = tf.cast(x_train, dtype=tf.float32)
y = Bernoulli(logits=(ed.dot(x, W) + b))
\end{lstlisting}

\subsubsection{Inference}

Here, we perform variational inference. Define the variational model to be a
fully factorized normal
\begin{lstlisting}
qW = Normal(mu=tf.Variable(tf.random_normal([D])),
            sigma=tf.nn.softplus(tf.Variable(tf.random_normal([D]))))
qb = Normal(mu=tf.Variable(tf.random_normal([1])),
            sigma=tf.nn.softplus(tf.Variable(tf.random_normal([1]))))
\end{lstlisting}

Run variational inference with the Kullback-Leibler divergence for $1000$ iterations.
\begin{lstlisting}
inference = ed.KLqp({W: qW, b: qb}, data={y: y_train})
inference.run(n_iter=1000, n_print=100, n_samples=5)
\end{lstlisting}
In this case
\texttt{KLqp} defaults to minimizing the
$\text{KL}(q\|p)$ divergence measure using the reparameterization
gradient.

\subsubsection{Criticism}

The first thing to look at are point-wise evaluations on the training dataset.

First form a plug-in estimate of the posterior predictive distribution.
\begin{lstlisting}
y_post = ed.copy(y, {W: qW.mean(), b: qb.mean()})
\end{lstlisting}

Then evaluate predictive accuracy
\begin{lstlisting}
print('Plugin estimate of posterior predictive log accuracy on training data:')
print(ed.evaluate('log_lik', data={x: x_train, y_post: y_train}))
> -3.12

print('Binary accuracy on training data:')
print(ed.evaluate('binary_accuracy', data={x: x_train, y_post: y_train}))
> 0.71
\end{lstlisting}

\begin{figure}[!htb]
\centering
\includegraphics[width=2.5in]{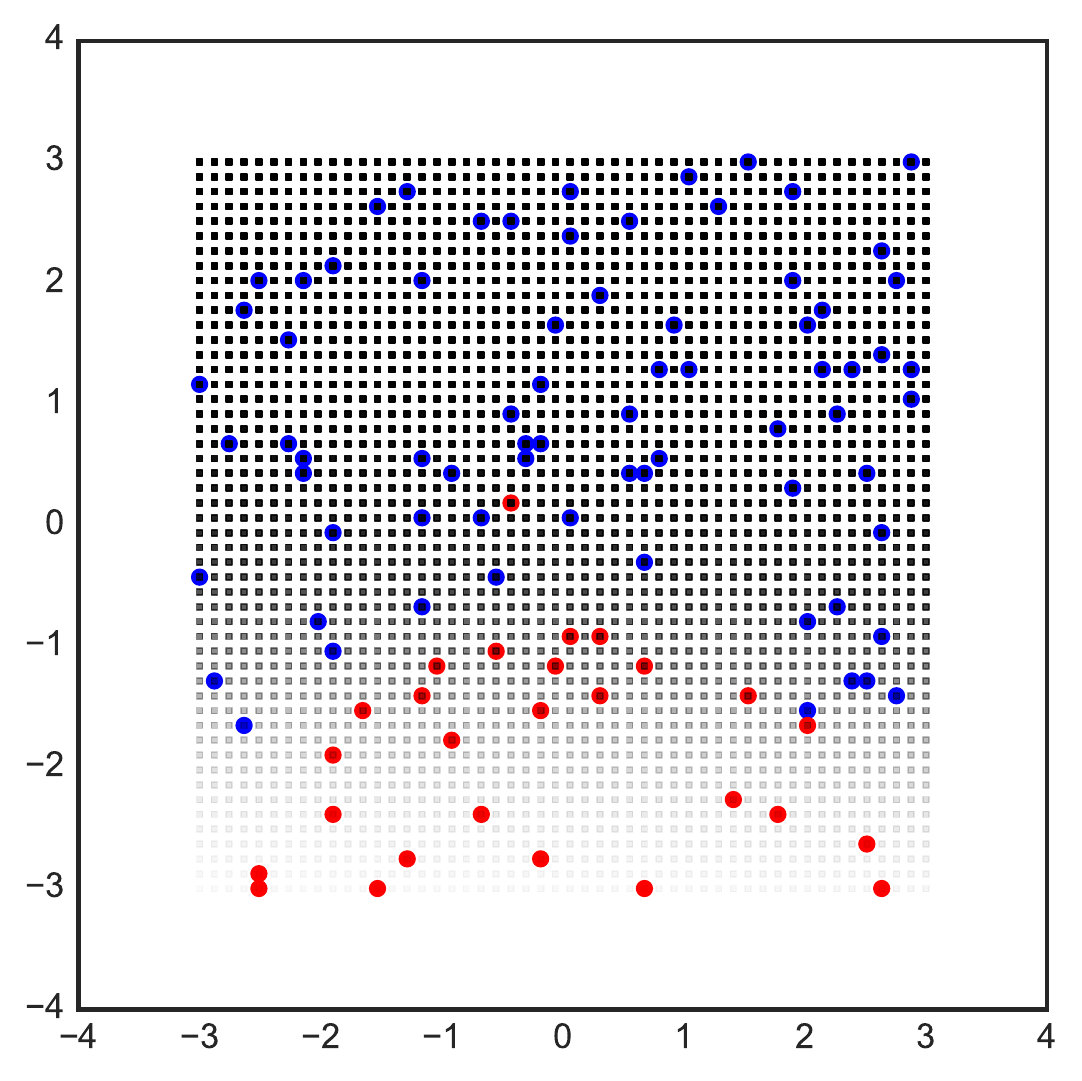}
\caption{Logistic regression struggles to separate the measurements.}
\label{fig:lr_linear}
\end{figure}

Figure\nobreakspace \ref {fig:lr_linear} shows the posterior label probability evaluated on a grid.
As expected, logistic regression attempts to fit a linear boundary between the
two label classes. Can a non-linear model do better?

\subsubsection{Model: Bayesian Neural Network Classification}

Consider parameterizing the label probability using a neural network; this
model is not limited to a linear relationship to the inputs $\mathbf{x}$, as in
logistic regression.

The model posits a likelihood for each observation $(\mathbf{x}_n,
y_n)$ as
\begin{align*}
\Pr(y_n=1)
  &=
  \text{logistic}
  \left(
  \text{NN}(\mathbf{x}_n \;;\; \mathbf{z})
  \right),
\end{align*}
where NN is a neural network and the latent random variable $\mathbf{z}$
contains its weights and biases.

We can specify a Bayesian neural network in Edward as follows. Here we specify a
fully connected two-layer network with two nodes in each layer; we posit
standard normal priors on all weights and biases.
\begin{lstlisting}
def neural_network(x, W_0, W_1, W_2, b_0, b_1, b_2):
    h = tf.nn.tanh(tf.matmul(x, W_0) + b_0)
    h = tf.nn.tanh(tf.matmul(h, W_1) + b_1)
    h = tf.matmul(h, W_2) + b_2
    return tf.reshape(h, [-1])

H = 2  # number of hidden units in each layer

W_0 = Normal(mu=tf.zeros([D, H]), sigma=tf.ones([D, H]))
W_1 = Normal(mu=tf.zeros([H, H]), sigma=tf.ones([H, H]))
W_2 = Normal(mu=tf.zeros([H, 1]), sigma=tf.ones([H, 1]))
b_0 = Normal(mu=tf.zeros(H), sigma=tf.ones(H))
b_1 = Normal(mu=tf.zeros(H), sigma=tf.ones(H))
b_2 = Normal(mu=tf.zeros(1), sigma=tf.ones(1))

x = tf.cast(x_train, dtype=tf.float32)
y = Bernoulli(logits=neural_network(x, W_0, W_1, W_2, b_0, b_1, b_2))
\end{lstlisting}

\subsubsection{Inference}

Similar to the above, we perform variational inference. Define the variational
model to be a fully factorized normal over all latent variables
\begin{lstlisting}
qW_0 = Normal(mu=tf.Variable(tf.random_normal([D, H])),
              sigma=tf.nn.softplus(tf.Variable(tf.random_normal([D, H]))))
qW_1 = Normal(mu=tf.Variable(tf.random_normal([H, H])),
              sigma=tf.nn.softplus(tf.Variable(tf.random_normal([H, H]))))
qW_2 = Normal(mu=tf.Variable(tf.random_normal([H, 1])),
              sigma=tf.nn.softplus(tf.Variable(tf.random_normal([H, 1]))))
qb_0 = Normal(mu=tf.Variable(tf.random_normal([H])),
              sigma=tf.nn.softplus(tf.Variable(tf.random_normal([H]))))
qb_1 = Normal(mu=tf.Variable(tf.random_normal([H])),
              sigma=tf.nn.softplus(tf.Variable(tf.random_normal([H]))))
qb_2 = Normal(mu=tf.Variable(tf.random_normal([1])),
              sigma=tf.nn.softplus(tf.Variable(tf.random_normal([1]))))
\end{lstlisting}

Run variational inference for $1000$ iterations.
\begin{lstlisting}
inference = ed.KLqp({W_0: qW_0, b_0: qb_0,
                     W_1: qW_1, b_1: qb_1,
                     W_2: qW_2, b_2: qb_2}, data={y: y_train})
inference.run(n_iter=1000, n_print=100, n_samples=5)
\end{lstlisting}

\subsubsection{Criticism}

Again, we form a plug-in estimate of the posterior predictive distribution.
\begin{lstlisting}
y_post = ed.copy(y, {W_0: qW_0.mean(), b_0: qb_0.mean(),
                     W_1: qW_1.mean(), b_1: qb_1.mean(),
                     W_2: qW_2.mean(), b_1: qb_2.mean()})
\end{lstlisting}

Both predictive accuracy metrics look better.
\begin{lstlisting}
print('Plugin estimate of posterior predictive log accuracy on training data:')
print(ed.evaluate('log_lik', data={x: x_train, y_post: y_train}))
> -0.170941

print('Binary accuracy on training data:')
print(ed.evaluate('binary_accuracy', data={x: x_train, y_post: y_train}))
> 0.81
\end{lstlisting}

\begin{figure}[!htb]
\centering
\includegraphics[width=2.5in]{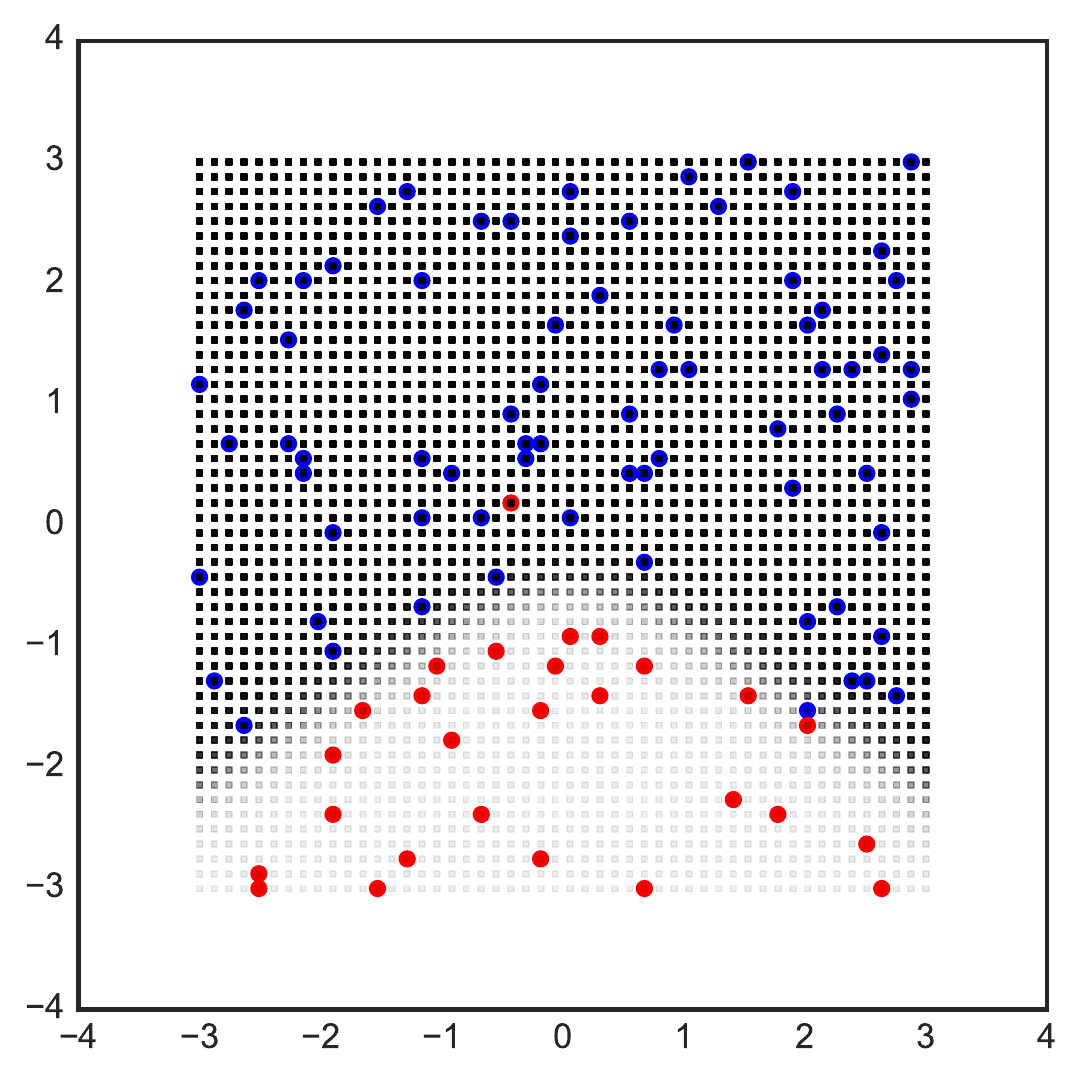}
\caption{A Bayesian neural network does a better job of separating the two
label classes.}
\label{fig:lr_nn}
\end{figure}

Figure\nobreakspace \ref{fig:lr_nn} shows the posterior label probability evaluated on a grid.
The neural network has captured the nonlinear decision boundary between the two
label classes.

\clearpage
\section{Acknowledgments}
Edward has benefited enormously from the helpful feedback and advice
of many individuals: Jaan Altosaar, Eugene Brevdo, Allison Chaney,
Joshua Dillon, Matthew Hoffman, Kevin Murphy, Rajesh Ranganath, Rif
Saurous, and other members of the Blei Lab, Google Brain, and Google
Research.
This work is supported by NSF IIS-1247664, ONR N00014-11-1-0651,
DARPA FA8750-14-2-0009, DARPA N66001-15-C-4032, Adobe, Google, NSERC
PGS-D, and the Sloan Foundation.

\clearpage
\bibliographystyle{apalike}
\bibliography{bib}

\end{document}